\documentclass{article}
\usepackage{authblk}
\usepackage[utf8]{inputenc}
\usepackage{geometry}
\usepackage{natbib}
\usepackage[colorlinks = true,
            linkcolor = blue,
            urlcolor  = blue,
            citecolor = black,
            anchorcolor = black]{hyperref}
\usepackage{enumitem,amssymb}
\usepackage{mdframed}
\usepackage{lastpage}
\usepackage{graphicx}
\usepackage{tabularx}
\usepackage[table,dvipsnames]{xcolor}

\usepackage{sectsty}
\usepackage{caption}
\usepackage{subcaption}
\definecolor{titlepagecolor}{RGB}{72,107,177}	
\definecolor{namecolor}{cmyk}{1,1,1,1} 
\usepackage{wrapfig}
\usepackage{comment}
\usepackage{longtable}

\usepackage{fancyhdr}
\usepackage{xpatch}
\xapptocmd{\footrule}{\color{black}}{}{}
\xpretocmd{\footrule}{\color{Mahogany!75}}{}{}
\xapptocmd{\footrule}{\color{black}}{}{}
\fancypagestyle{titlepage}{%
  \fancyhf{} 
  \color{white}
  \vspace{3cm}
  \fancyfoot[L]{\includegraphics[width=1cm]{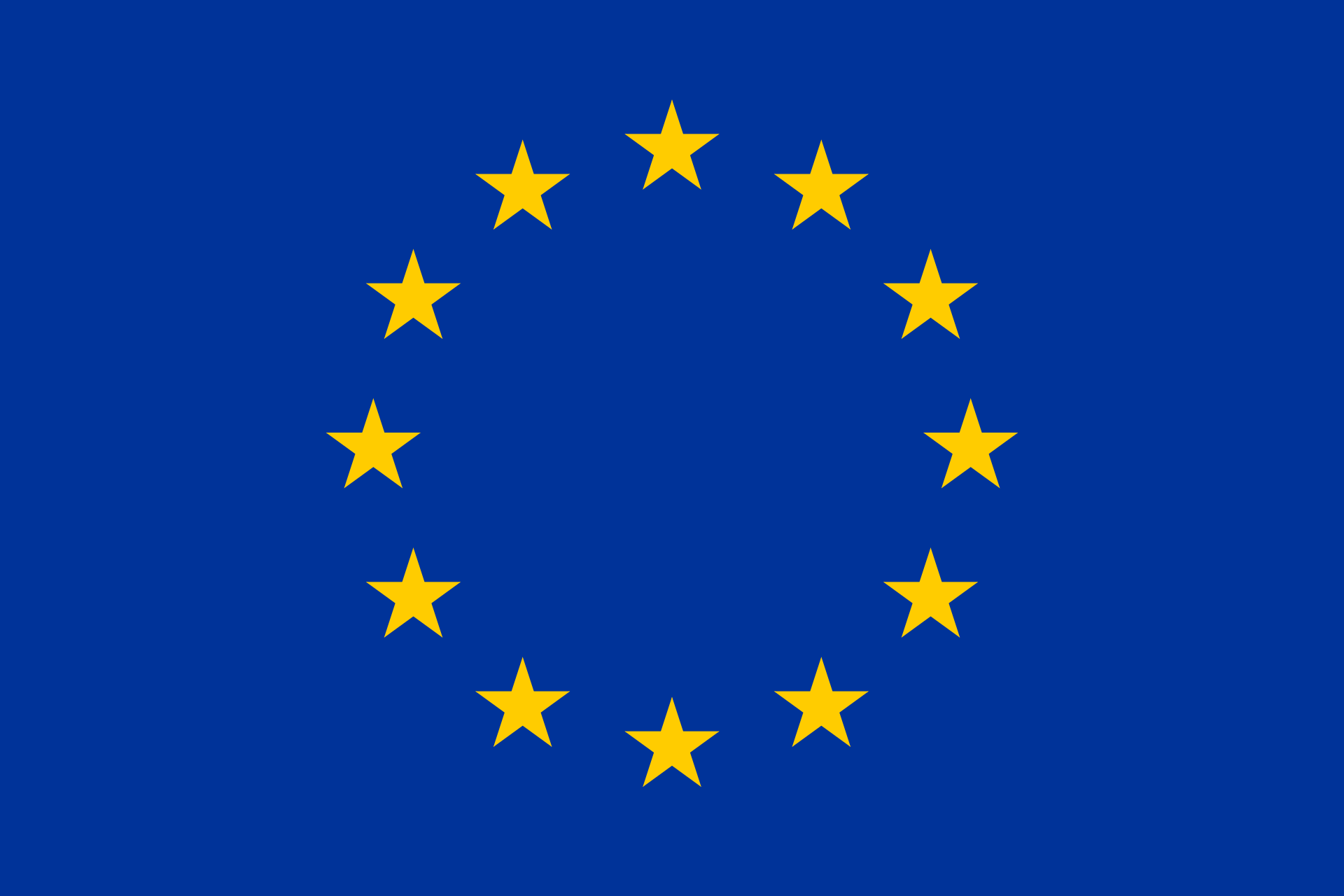}{\raisebox{7pt}{\footnotesize{\textcolor{white}{ Horizon 2020 Grant agreement No. 951815 }}}}}
  \color{Mahogany!75}
  
}
\usepackage{adjustbox}
\usepackage{array}

\newlist{checkboxes}{itemize}{2}
\setlist[checkboxes]{label=$\square$}
\usepackage{pifont}

\ExplSyntaxOn
\NewDocumentCommand{\fillwithspace}{m}{\prg_replicate:nn {\fp_to_int:n { #1/\dim_to_fp:n {\baselineskip}}}{\mbox{}\par}}
\ExplSyntaxOff

\usepackage{amsmath}
\usepackage{tikz}
\usepackage[scaled]{helvet}
 
\usepackage[OT1]{fontenc}
\usepackage{makecell}
\usepackage{mathastext}

\usepackage{booktabs} %
\usepackage{xparse}   %

\newcommand{\submm}{\mbox{(sub-)mm}}
\newcommand{\Submm}{\mbox{(Sub-)mm}}

\def\M87{M87$^*$}
\def\m87{M87$^*$}

\begin{document}

\begingroup
\hypersetup{hidelinks}
\begin{titlepage}
\thispagestyle{titlepage}

\newgeometry{left=4.5cm} %
\pagecolor{titlepagecolor}
\noindent
\includegraphics[width=6cm]{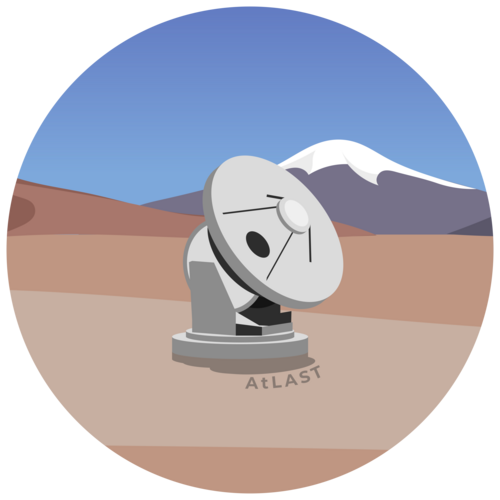}\\[-1em]
\color{white}
\colorbox{Mahogany!75}{\makebox[25pt][l]{\rule{1.3\textwidth}{1pt}}}

\noindent
\begin{sloppypar}
\textbf{\textsf{\href{https://orcid.org/0000-0001-8568-6336}{Mark Booth}$^1$, 
\href{https://orcid.org/0000-0001-9443-0463}{Pamela Klaassen$^1$}, 
\href{https://orcid.org/0000-0003-0522-6941}{Claudia Cicone}$^2$, 
\href{https://orcid.org/0000-0003-3816-5372}{Tony Mroczkowski}$^3$, 
\href{}{Martin A. Cordiner}$^4$, 
\href{https://orcid.org/0000-0003-3586-4485}{Luca Di Mascolo}$^{5,6,7,8}$, 
\href{https://orcid.org/0000-0002-6773-459X}{Doug Johnstone}$^{9,10}$, 
\href{https://orcid.org/0000-0002-6327-5154}{Eelco van Kampen}$^3$, 
\href{https://orcid.org/0000-0002-2419-3068}{Minju M. Lee}$^{11,12}$, 
\href{https://orcid.org/0000-0001-9773-7479}{Daizhong Liu}$^{13,14}$, 
\href{https://orcid.org/0000-0003-1842-8104}{John Orlowski-Scherer}$^{15}$, 
\href{https://orcid.org/0000-0003-4357-3450}{Amélie Saintonge}$^{16,17}$, 
\href{https://orcid.org/0000-0002-3532-6970}{Matthew W. L. Smith}$^{18}$, 
\href{https://orcid.org/0000-0002-8178-1042}{Alexander Thelen}$^{19}$, 
Sven Wedemeyer$^{2,20}$, 
\href{https://orcid.org/0000-0002-9475-4254}{Kazunori Akiyama}$^{21,22,23}$,
\href{https://orcid.org/0000-0002-2041-8784}{Stefano Andreon}$^{24}$, 
\href{https://orcid.org/0000-0002-1959-7201}{Doris Arzoumanian}$^{25}$,
\href{https://orcid.org/0000-0002-5268-2221}{Tom J. L. C. Bakx}$^{26}$,
\href{https://orcid.org/0000-0001-6118-2985}{Caroline Bot}$^{27}$, 
\href{https://orcid.org/0000-0003-4056-9982}{Geoffrey Bower}$^{28}$,
\href{}{Roman Brajša}$^{29}$, 
\href{https://orcid.org/0000-0002-3805-0789}{Chian-Chou Chen}$^{30}$, 
\href{https://orcid.org/0000-0001-9759-4797}{Elisabete da Cunha}$^{31}$, 
\href{https://orcid.org/0000-0002-5881-3229}{David Eden}$^{32}$, 
\href{https://orcid.org/0000-0003-4117-8617}{Stefano Ettori}$^{33,34}$, 
\href{https://orcid.org/0000-0003-4224-6829}{Brandt Gaches}$^{26}$, 
\href{https://orcid.org/0000-0003-0917-9636}{Evanthia Hatziminaoglou}$^{3,35,36}$, 
\href{https://orcid.org/0000-0002-1018-6203}{Patricia Luppe}$^{37}$, 
\href{https://orcid.org/0000-0002-6777-6490}{Benjamin Magnelli}$^{38}$, 
\href{https://orcid.org/0000-0001-6208-1801}{Jonathan P. Marshall}$^{30}$, 
\href{https://orcid.org/0000-0002-7430-3771}{Francisco Miguel Montenegro-Montes}$^{39}$, 
Michael Niemack$^{40}$,
\href{https://orcid.org/0000-0001-9540-9121}{Conor Nixon}$^{41}$, 
\href{https://orcid.org/0000-0002-4278-3168}{Imke de Pater}$^{42}$, 
\href{https://orcid.org/0000-0002-6255-8240}{Yvette Perrott}$^{43}$, 
\href{https://orcid.org/0000-0002-6248-398X}{Sandra I. Raimundo}$^{44,45}$, 
\href{https://orcid.org/0000-0002-0528-8125}{Elena Redaelli}$^{46}$, 
\href{https://orcid.org/0000-0002-3880-2450}{Anita Richards}$^{47}$,
\href{https://orcid.org/0000-0002-1383-0746}{Matus Rybak}$^{48,49,50}$, 
\href{https://orcid.org/0000-0001-7301-6415}{Nikolina Šarčević}$^{51}$, 
\href{https://orcid.org/0000-0002-3913-7114}{Dmitry Semenov}$^{52}$, 
\href{https://orcid.org/0000-0002-6787-5245}{Silvia Spezzano}$^{46}$, 
\href{https://orcid.org/0000-0002-2996-305X}{Sundar Srinivasan}$^{53}$, 
\href{https://orcid.org/0000-0002-5812-9232}{Thomas Stanke}$^{46}$, 
\href{https://orcid.org/0000-0001-9493-0169}{Paola Andreani}$^{3}$, 
\href{https://orcid.org/0000-0003-3315-5626}{Maria T. Beltrán}$^{54}$, 
\href{https://orcid.org/0000-0002-5344-820X}{Bryan J. Butler}$^{55}$, 
\href{https://orcid.org/0000-0001-5804-1428}{Sebastiano Cantalupo}$^{56}$, 
\href{https://orcid.org/0000-0003-0407-8115}{Miguel Chavez Dagostino}$^{57}$, 
\href{https://orcid.org/0000-0002-5259-4774}{Ana Duarte-Cabral}$^{18}$, 
\href{https://orcid.org/0000-0003-2983-815X}{Bjorn Emonts}$^{58}$, 
\href{https://orcid.org/0000-0001-5834-9588}{Leigh Fletcher}$^{59}$, 
\href{https://orcid.org/0000-0003-2520-8396}{Dale E. Gary}$^{60}$, 
\href{https://orcid.org/0000-0003-3889-2609}{Stanislav Gunar}$^{61}$, 
\href{https://orcid.org/0000-0001-5397-6961}{Alvaro Hacar}$^{62}$, 
\href{https://orcid.org/0009-0003-9551-4089}{Bendix Hagedorn}$^{2}$, 
\href{https://orcid.org/0000-0001-8541-8024}{Tomek Kaminski}$^{63}$,
Fiona Kirton,
\href{https://orcid.org/0000-0002-9068-3428}{Katherine de Kleer}$^{64}$, 
\href{https://orcid.org/0000-0002-8078-0902}{Eduard Kontar}$^{65}$, 
\href{https://orcid.org/0000-0002-4336-0730}{Yi-Jehng Kuan}$^{66,30}$,
John Lightfoot,
\href{https://orcid.org/0000-0001-5357-6538}{Enrique Lopez-Rodriguez}$^{67}$, 
\href{https://orcid.org/0000-0001-6795-5301}{Andreas Lundgren}$^{68}$, 
\href{https://orcid.org/0000-0001-7694-4129}{Stefanie N. Milam}$^{4}$, 
\href{https://orcid.org/0000-0002-1571-7931}{Atul Mohan}$^{69,70}$, 
\href{https://orcid.org/0000-0002-9171-2702}{Raphael Moreno}$^{71}$, 
\href{https://orcid.org/0000-0001-7856-084X}{Galina G. Motorina}$^{72,73}$, 
\href{https://orcid.org/0000-0002-9820-1032}{Arielle Moullet}$^{58}$, 
\href{https://orcid.org/0000-0002-8557-3582}{Kate Pattle}$^{16}$, 
\href{https://orcid.org/0000-0002-4590-0040}{Alberto Pellizzoni}$^{74}$, 
\href{}{Nicolas Peretto}$^{18}$, 
Joanna Ramasawmy,
\href{https://orcid.org/0000-0001-5231-2645}{Claudio Ricci}$^{75,76}$,
\href{https://orcid.org/0000-0002-3351-2200}{Andrew J. Rigby}$^{77}$, 
\href{https://orcid.org/0000-0002-3078-9482}{\'Alvaro S\'anchez-Monge}$^{78,79}$, 
\href{https://orcid.org/0000-0001-7353-9101}{Maryam Saberi}$^{20}$, 
\href{https://orcid.org/0000-0002-2350-3749}{Masumi Shimojo}$^{80,81}$, 
\href{https://orcid.org/0000-0002-9714-3862}{Aurora Simionescu}$^{50}$, 
\href{https://orcid.org/0000-0001-5392-909X}{Mark Thompson}$^{77}$, 
\href{https://orcid.org/0000-0003-1665-6402}{Alessio Traficante}$^{82}$, 
\href{https://orcid.org/0000-0002-8853-9611}{Cristian Vignali}$^{83,33}$, 
\href{https://orcid.org/0000-0002-8574-8629}{Stephen M. White}$^{84}$
}}
\textcolor{namecolor}{\textsf{\\See Appendix \ref{app:affil} for the author affiliations.}}

\end{sloppypar}
\vfill
\noindent
{\huge \textsf{AtLAST Science Overview Report}}

\noindent
\textsf{June, 2024}
\end{titlepage}
\restoregeometry %
\nopagecolor%
\color{black} %

\clearpage

\tableofcontents
\endgroup
\clearpage

\section*{Executive Summary}
\addcontentsline{toc}{section}{\protect\numberline{}Executive Summary}
\color{black}

Submillimeter and millimeter wavelengths provide a unique view of the Universe, from the gas and dust that fills and surrounds galaxies to the chromosphere of our own Sun. Current single-dish facilities have presented a tantalising view of the brightest (sub-)mm sources, and interferometers have provided the exquisite resolution necessary to analyse the details in small fields, but there are still many open questions that cannot be answered with current facilities.  Using sub-mm wavelengths to observe the universe allows us to unlock the motions of gas and dust throughout its lifecycle. Many of these populations are faint and diffuse, and so missed by current observatories. In order to make major advances in (sub-)mm astronomy, what is needed now is a facility capable of rapidly mapping the sky spatially, spectrally, and temporally, which can only be done by a high throughput, single-dish observatory. This is driving the design of the Atacama Large Aperture Submillimeter Telescope (AtLAST).

In the distant universe, the goal is to study galaxy populations as a whole, using a number of techniques (e.g. SZ, line intensity mapping) to enhance our understanding of the processes that shaped these galaxies and shapes their evolution. With AtLAST we will finally be able to probe the size and sensitivity regime required to find `normal' galaxies at high-redshift.  The circumgalactic medium, at all redshifts, is showing itself to be a faint but key ingredient in that evolution, acting as a bridge between the interstellar and intergalactic media enabling the flow of fresh materials between them.  At lower redshifts, in the nearby universe, we are able to start resolving the processes of star and planet formation embedded within the galaxies that started forming much earlier in the evolution of the Universe. With a better understanding of the magnetic field structures, the heating, cooling, chemistry and dust composition of nearby galaxies, including the Magellanic Clouds and the Milky Way (the best resolved structures to study), we can quantify the balance of forces on smaller (molecular cloud) scales that then drive the evolution on galaxy and cluster scales. Closer in, high cadence observations of the Solar chromopshere will further our understanding of space weather.  We will be able to resolve the planets in our solar system, study their weather patterns and, in some cases, even resolve their moons.  Studies of the chemistry of comets will allow us to probe the origins of Earth's water.

A single-dish facility with a 50~m diameter provides the sensitivity to trace the gas and dust from protostellar to circumgalactic scales (both near and far), and the resolution to spatially resolve weather patterns on the planets and moons of our solar system. This level of sensitivity and resolution will enable the systematic detection of the relativistic and kinetic SZ effects and, for the first time, the emission from `normal' galaxies at high redshift, which is  hidden below the confusion limits of current and planned single-dish observatories.

The large field of view proposed for AtLAST ($\geq$1 deg$^2$), will enable the large format (multiplexed) instrumentation that will be able to detect structures at these scales. The benefits are two fold here: detecting emission on larger scales, and increased survey efficiency.  The sensitivity to large scale structure these instruments provide comes from the requirement to take background measurements at much larger offsets instead of within the instantaneous field of view of the instrument.  The ability to fill the large field of view with highly multiplexed instruments increases mapping efficiency and serendipity.  There are a number of large scale survey science cases described here (i.e. the Galactic plane, the Magellanic clouds and nearby galaxies) that will require 1000s of times fewer pointings if the AtLAST FoV were filled when compared to current generation facilities. With this large FoV,
efficient transient/variability detection algorithms built into the data processing algorithms will enable a wide variety of serendipitous detections. 

Similarly, high bandwidth receivers (in both  the continuum and spectral domain) will enable rapid surveys by reaching higher sensitivities / detecting more lines in a given observation than current generation facilities. This exploit is similar to the reasoning for the ALMA WSU. As receiver technology advances over the next decade, polarisation preservation is becoming in-built, and not only will these highly sensitive receiver technologies study the spatial distribution of gas and dust, but its magnetic properties as well.

The high sensitivity provided by AtLAST will result in deep observations on short timescales,
which enables cadenced observations to be performed on various timescales: from a few seconds, to hours, to years depending on the science need (i.e. solar flares, cloud motions in planetary atmospheres, protostellar flares, AGNs and GRBs).

This study has uncovered a number of key science goals that cannot be tackled with current generation facilities, even if their instrumentation were upgraded. To satisfy the science needs of the community an entirely new facility is required with high throughput (large diameter and field of view) and the high sensitivity that comes with it. The lifecycle of gas and dust in the Universe is proving to not be a simple, steady state process, but one full of surprises. AtLAST will be able to unlock hidden populations of objects, study how they interact with their environments, and how they vary over time.

One of the key deliverables for the science work package of the AtLAST Horizon 2020 design study is a robust set of science-driven requirements for the telescope design. In this regard, we have solicited input from the astronomy community, organised into science working groups (SWGs) covering different themes. These SWGs have produced a collection of papers describing the key science drivers in each area and the requirements they place on the observatory\footnote{These have all been submitted to the journal Open Research Europe: \url{https://open-research-europe.ec.europa.eu/collections/atlast/about}.}. The purpose of this document is to summarise the results of these papers and present the combined requirements set by them.

\begin{table}[hbt]
    \centering
    \begin{tabular}{p{1.6cm}p{4.5cm}p{4.5cm}p{4.5cm}}

& {\cellcolor{RoyalBlue!50} ~}  
  \textit{\textbf{Where are all the baryons?}} 
& {\cellcolor{RoyalBlue!50} ~}
  \textit{\textbf{How do structures interact with their environments?}}
& {\cellcolor{RoyalBlue!50} ~}
  \textit{\textbf{What does the time-varying (sub-)mm sky look like?}}\\ 
  \hline
 
{\cellcolor{Mahogany!75} \color{white}\bf Detailed science goal} 

&  Measuring the total gas and dust content of the Milky Way and other galaxies, in the interstellar, circumgalactic, and intergalactic media, reaching down to the sensitivities required to probe the typical populations of sub-mm sources.

&  Understanding the lifecycle of gas and dust near and far; mapping the baryon cycle on multiple-scales; observing the interplay between gravity, radiation, turbulence, magnetic fields, and chemistry and their mutual feedback.  

& Identifying the mechanisms responsible for time variability across astrophysical sources: from the Solar corona and other objects in our solar system to luminosity bursts in everything from protostars to active galactic nuclei.\\
\hline

{\cellcolor{Mahogany!75} \color{white}\bf Detailed technical specification}

&  \textbf{\textit{High sensitivity to the faint signals}} (at sub-mK levels)  \textit{\textbf{on large scales}} ($\geq$1 deg$^2$) from even the most diffuse and cold gas through sub-mm line tracers. Wide field ($>$500 deg$^2$)  continuum surveys capturing the plane of our galaxy and resolving 80\% of the cosmic infrared background, probing typical populations and looking back over 90\% of the age of the Universe.   

& \textit{\textbf{High spectral resolution}} and \textit{\textbf{polarisation }}measurements on the relevant size scales for cores (0.1 pc, our galaxy), clumps (10 pc, nearby galaxies) and cloud complexes ($\sim$ few kpc, distant universe) to quantify the chemistry, disentangle the dynamics, and measure the magnetic fields working together to shape the evolution of structures within their larger-scale environments. 

& An operations model that allows for \textit{\textbf{highly cadenced and rapid response observations}} and data reduction pipelines with in-built \textit{\textbf{transient detection algorithms}}\textbf{; }high time-resolution (few seconds) observations of our Sun and other stars.\\
\multicolumn{4}{c}{}\\
    \end{tabular}
    \vspace{-36pt}
    \caption{Key Science Drivers for AtLAST}
    \label{tab:key_questions}
\end{table}

From this process we find that there are three key themes where AtLAST will provide transformational science, which are described in Table \ref{tab:key_questions}.
To accomplish this, AtLAST will require high sensitivity, high angular resolution, continuum and spectral line instrumentation which preserves polarisation, solar observing capabilities and a highly flexible scheduling model capable of snapshot observations, large surveys, rapid response to transient triggers and a transient pipeline capable of sending transient triggers.

With a telescope design and operations model capable of fulfilling these requirements and an innovative sustainability plan, AtLAST will be capable of serving the astronomy community for decades to come.

\clearpage

\section{Introduction}
\color{black}
\subsection{Sub-mm astronomy in the context of past, present and future \submm{} facilities}
Observations in the sub-mm and mm wavelength regime (by which we are roughly referring to wavelengths between 0.3 and 10~mm) provide the opportunity to view a wide range of astrophysical phenomena, from cold dust and gas to high energy sources. Dust at 10~K has a peak wavelength of around 300~$\mu$m and carbon monoxide (one of the most common molecules in the Universe) has many emission lines at (sub-)mm wavelengths. These are all redshifted to longer wavelengths the more distant the galaxy. They provide ways to measure the dust and gas mass of galaxies, their circumgalactic media (CGM) and clouds, cores and circumstellar discs within our own galaxy. Chemical signatures allow us to study not just the chemistry of astrophysical sources but also their dynamics. Hot gas can also be studied in the \submm{} due to the scattering of CMB photons by free electrons in the gas, which can be detected in galaxy clusters, protoclusters and even around individual galaxies. \Submm{} observations are even possible of the brightest source in the sky -- our own Sun.

\begin{figure*}[hb!]
    \centering
    \includegraphics[width=0.7\textwidth]{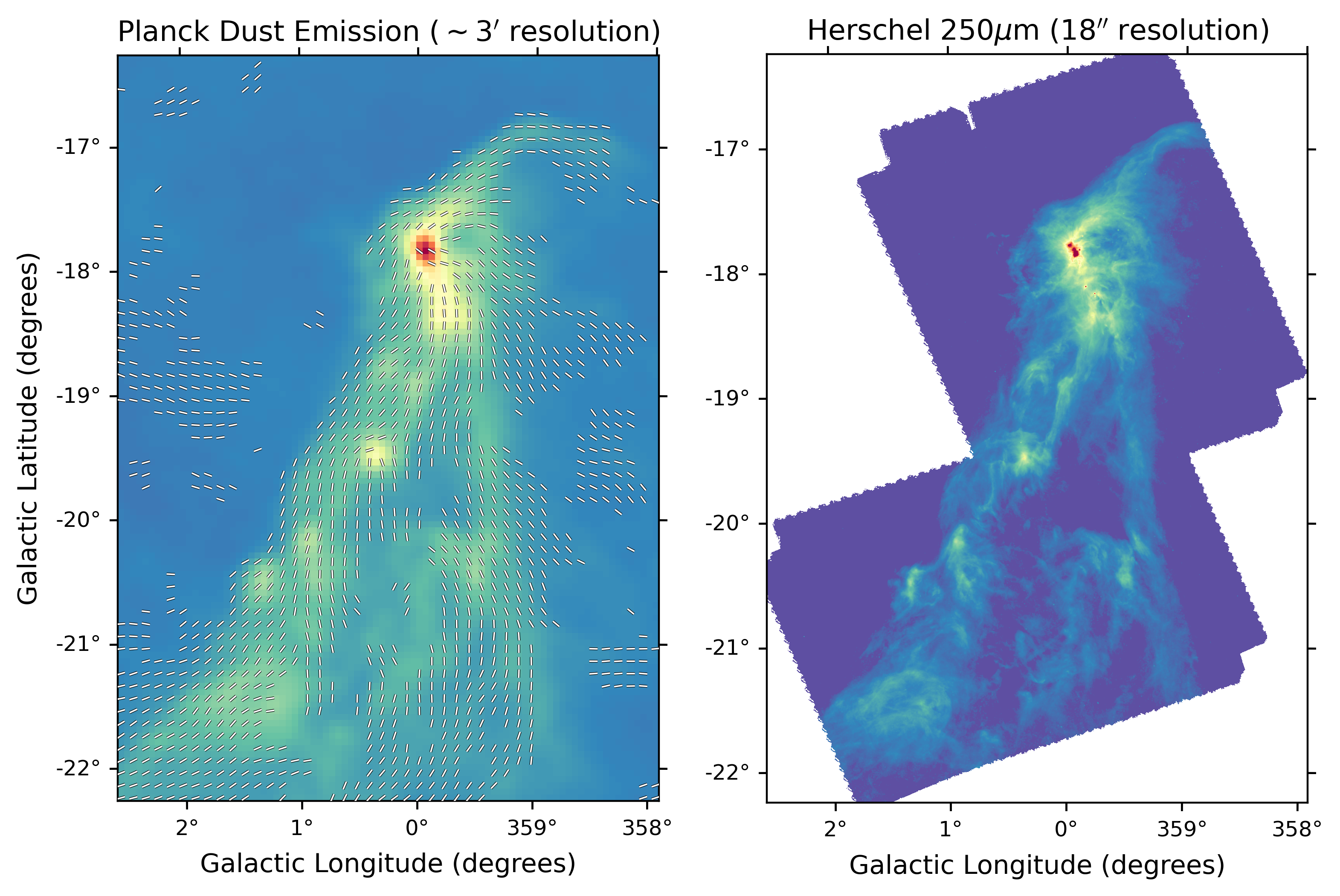}
    \caption{Comparison of Planck 353\,GHz (\citealt{planck2015_intXIX}; left) and Herschel 1.2\,THz (250\,$\mu$m; \citealt{bresnahan2018}; right) mapping of the Corona Australis molecular cloud.  While the Planck data shows the overall shape of the magnetic field in the cloud, it cannot resolve individual sites of star formation.  \textbf{With AtLAST, we will map magnetic fields across the entire Galactic plane with a resolution at 345\,GHz that is four times better than that shown in the right-hand panel, thereby resolving individual star-forming cores within molecular clouds.} \citep[Figure reproduced from][]{Klaassen2024}}
    \label{fig:cra}
\end{figure*}

AtLAST builds upon 60 years of sub-mm astronomy. In the 1970s and 80s the sub-mm sky first opened up to astronomers through dedicated facilities such as the Kuiper Airborne Observatory (KAO) and the James Clerk Maxwell Telescope (JCMT), based on the ability to push far-infrared technologies to longer wavelengths while simultaneously pushing radio and mm-wave technologies to shorter ones. Their pioneering efforts led to the discovery of sources as close as the rings of Uranus \citep{Elliot1977}, and as far away as entirely new populations of galaxies - first dubbed `sub-mm galaxies' because they had never been detected at other wavelengths \citep{Blain2002}. Combining the JCMT and the Caltech Sub-mm Observatory (CSO) into a sub-mm interferometer \citep{Carlstrom1994} then also paved the way for sub-mm interferometers like the SMA, NOEMA and ALMA to be seen as technically feasible.

Since those early days, and as our understanding of the sub-mm universe has grown, the interest in observing it has blossomed - to the point of ALMA routinely being the most oversubscribed ground based observatory in the world. Hand in hand with this scientific curiosity has been the technological development required to obtain those science goals. 

At sub-mm wavelengths, we are sensitive to gas and dust across cosmic time. We often observe this through the thermal emission of cold dust and low energy transitions of molecules. But, at these wavelengths, we can also probe atomic, forbidden and ionised gas lines, the warm/hot ionised medium (WHIM) of galaxies and even the magnetic fields permeating these gas and dust populations (see Figure \ref{fig:cra} for an example). Some of the signals we detect emit on small scales (of order arcsec and smaller) but some emit on very large scales (few-10s of degrees) as shown in the rest of this document.

\begin{wrapfigure}{R}{0.45\textwidth} %
    \centering
    \includegraphics[width=0.4\textwidth]{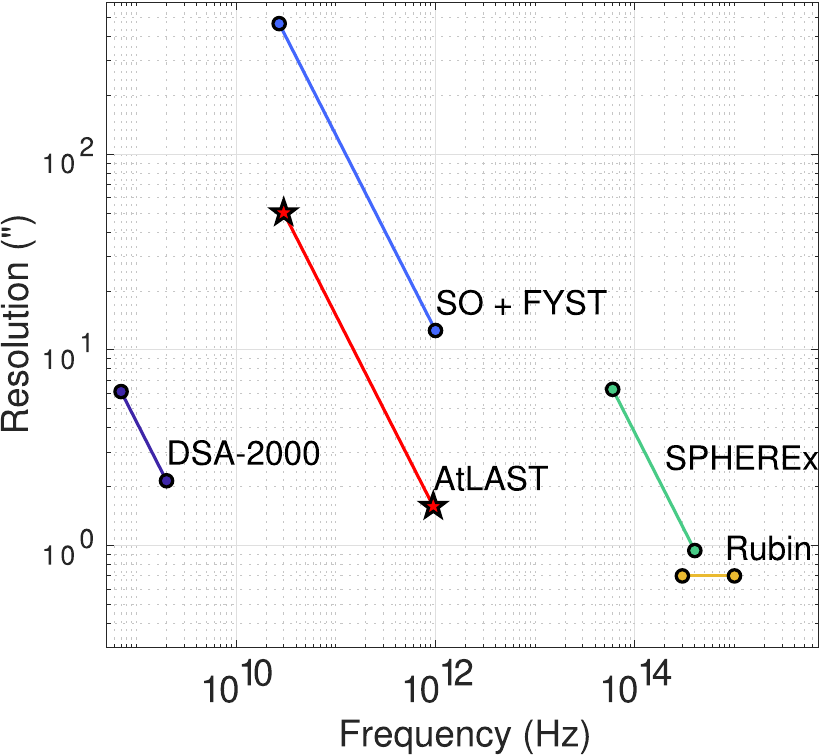}
    \caption[] 
    {Angular resolution versus observing frequency for AtLAST and current and planned telescopes with fields of view $\geq$1~deg$^2$.} %
    \label{fig:largeaperture}
\end{wrapfigure}

We are now a long way from the single pixel continuum and spectroscopic cameras that first populated those early observatories, with single-dish facilities like APEX, ASTE, the IRAM-30m, and upgrades to the CSO (now decomissioned) and JCMT hosting continuum cameras with 100s to 1000s of pixels to detect dust continuum and polarisation, 10s of pixel spectroscopic instruments able to resolve gas kinematics with precisions below 0.01 km/s (R $>$ 10$^8$), and starting to deploy low spectral resolution integral field units (IFUs) with broad bandwidths aimed at determining the redshifts of all objects in their fields of view.  The above are general purpose observatories with broad suites of instruments adaptable to the changing needs of their diverse science communities.   In parallel to this, and on the horizon are a number of 6-12 m class observatories dedicated to specific science goals/surveys.

Single dish observatories have shaped our understanding of the sub-mm sky, and have found some of the most interesting objects for interferometric followup. As we learn more and more about the sub-mm sky through single dish observations there are some questions that can only be answered with higher spatial resolution observations, which requires the use of interferometers. However, interferometers trade high angular resolution for their ability to detect large scale structures due to the lack of short baselines. This is a trade-off worth making when the structures being observed do not emit on large scales, information about large scale  structure is not important, or that information can be folded in latter using dedicated total power telescopes. This latter technique requires combining data from observatories like the JCMT and SMA, IRAM-30m and PdBI/NOEMA, or within a single observatory as is done in ALMA: combining the 12~m array with its compact array and total power antennas.  Single dish observatories are powerful observatories in their own right, but can also improve the results of interferometers by detecting the large scale emission that interferometers are blind to, yet the current observatories are limited in this regard as they lack the sensitivity and dish size necessary to complement ALMA, particularly for faint, extended sources \citep{Plunkett+23}.

Looking ahead to the 2030s and beyond, the landscape of long wavelength astronomy is set to change again. With interferometric facilities like the ALMA Wideband Spectroscopic Upgrade \citep[WSU;][]{Carpenter+2023} and SKAO\footnote{\url{https://skao.int}}  coming online at the beginning of that timeframe and potentially the ngVLA\footnote{\url{https://ngvla.nrao.edu}} along with it in combination with dedicated large area survey telescopes like SO, CMB-S4 and FYST, we are entering a new golden age for unlocking the secrets of the long wavelength sky. In addition to the main science drivers of the latter set of single dish facilities, they will provide finder charts for followup observations with these next generation interferometers.  But these surveys will be inherently shallow (i.e. limited to 0.1 - 1 mJy over 100 - 600 GHz)  because of their limited collecting areas and high confusion limits dictated by their  6-10~m individual apertures 
 \citep[see, e.g., Figure 13 of][]{Blain2002}. 
A 50~m class single-dish facility will be able to probe deeper (with confusion limits 10x lower at 100 GHz, and 1000x lower at 600 GHz), at higher spatial resolution (3x the linear resolution of JCMT, 4x that of APEX, and 8x that of the upcoming 6~m facilities -- see Figure \ref{fig:largeaperture}) and much more quickly (single point sensitivities 9x, 16x and 64x those of the JCMT, APEX and upcoming 6~m class facilities per unit integration time) than other upcoming observatories. This combination of capabilities means with a 50~m class sub-mm telescope that has a large field of view, we can push to the `normal' population of galaxies at high redshift (well below the confusion limits of current and upcoming facilities), see the large scale, but diffuse and faint, interstellar and circumgalactic media of galaxies including our own, resolve variability across the nearby universe and weather patterns on planets and moons in our solar system, while ensuring that the next generation interferometers will have many more interesting objects to observe over their lifetimes (i.e. they will not become `source-starved'), while putting those objects into their environmental context.

\subsection{The AtLAST Design Study, and purpose and scope of this document}

The AtLAST design study is a 3.5 year Horizon 2020 funded project with the aim of deriving a next generation single-dish sub-mm telescope designed with science and sustainability at the forefront \citep{Klaassen+20}. The project spans six work packages including (WP1) overall management, (WP2) telescope design, (WP3) site selection, (WP4) telescope operations, (WP5) sustainability studies and (WP6) science. Since the beginning of the project, a detailed optical and telescope design has been developed \citep{Gallardo2024, Mroczkowski2024,Kiselev2024,Puddu2024,Reichert2024}, two sites on the Atacama plateau have been monitored and studied for suitability\footnote{\url{https://www.atlast.uio.no/design-study/wp3-site/d3.1_site_selection_criteria.pdf}}, and an operations plan taking equality, diversity and inclusion principles into account has taken shape.  Research into powering the telescope using sustainable methods has been conducted \citep{Viole2023,Viole2024b, Viole2024}, including consultations with the local San Pedro de Atacama and astronomical communities to ensure their views on the future of sustainable astronomy are heard and understood \citep{Valenzuela-Venegas2023}. Refining the science case is the work package discussed in this report.

The purpose of this document is to summarise the work of the science community over the Horizon 2020 AtLAST design study to derive and refine the science drivers and understand what we could achieve with such a disruptive and innovative leap in technology.  In Section \ref{sec:themes} we summarise the science drivers for the telescope. In Section \ref{sec:requirements} we describe the constraints placed on the telescope by the many and varied science drivers and summarise this in a matrix of requirements.

\newpage

\section{Science Themes}
\color{black}

\label{sec:themes}

\begin{figure}[hbt]
    \centering
    \includegraphics[width=0.9\textwidth]{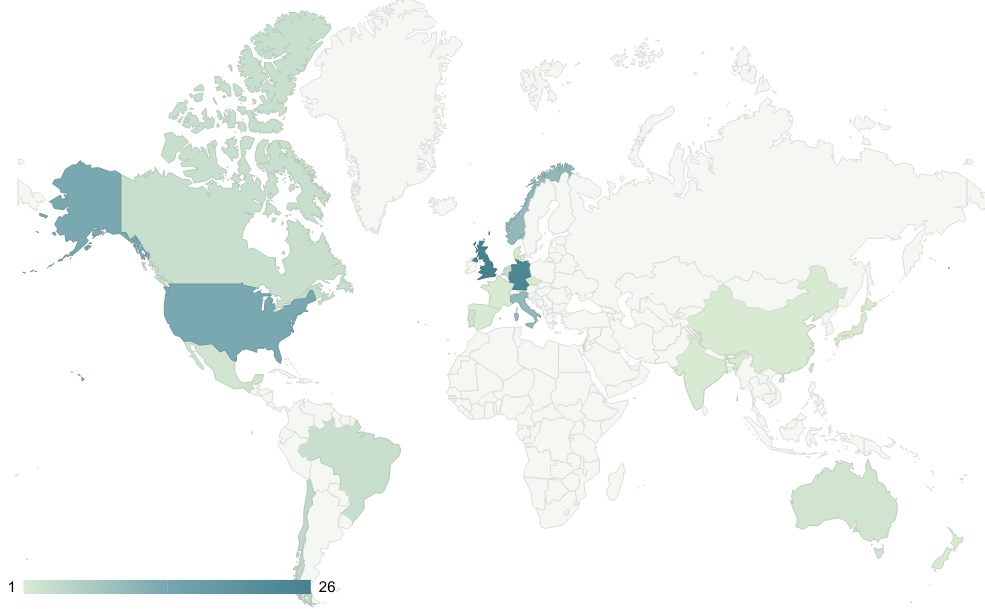}
    \caption{Map showing the contributions to the initial use case submission part of the community consultation for defining the key science drivers for AtLAST. Figure reproduced from \cite{Ramasawmy2022}.}
    \label{fig:map}
\end{figure}

The science consultation for the AtLAST design study began with polling the worldwide community to submit individual science use cases. This resulted in the participation of over 100 scientists from 22 countries submitting 28 use cases (see figure \ref{fig:map} for their geographic distribution). These use cases saw the community requesting both spectroscopy and imaging, and everything from limited area mapping to large scale surveys. From these submissions, four key science categories were identified:  The Sun and Solar System, the Milky Way, Nearby Galaxies, and the Distant Universe and Cosmology. Embedded in these themes was also the need for cadenced (and target of opportunity) observations to measure variability in the sub-mm sky. Together these use cases showed a great desire from the community for a new single dish facility with higher spatial resolution, greater mapping efficiency, orders of magnitude increases in sensitivity,  and highly multiplex instruments being able to observe continuum and line emission, often requiring polarisation preservation.

Based on these themes, working groups were created to do more in-depth studies into the feasibility of answering new questions in these subject areas. Where gaps were identified in the initial submissions, under-represented science communities were targeted for more participation to ensure the derived science cases were as well rounded as possible. Alongside this process, a sensitivity calculator was developed to allow the working groups to quantify their science goals with observing time estimates. This calculator is described in more detail in Appendix \ref{app:senscalc}. These working groups met periodically throughout the design study and the results of their investigations are captured in a series of white papers \citep{vanKampen2024, DiMascolo2024,Lee2024, Liu2024,Klaassen2024,Cordiner2024,Wedemeyer2024, Orlowski-Scherer2024}. Below, we summarise the work captured in these white papers along with some highlights of the science presented in the original use case studies \cite{Ramasawmy2022} and related works. 

\begin{wrapfigure}[40]{R}{0.42\textwidth}
\vspace{-72pt}
\centering
    \includegraphics[width=0.4\textwidth]{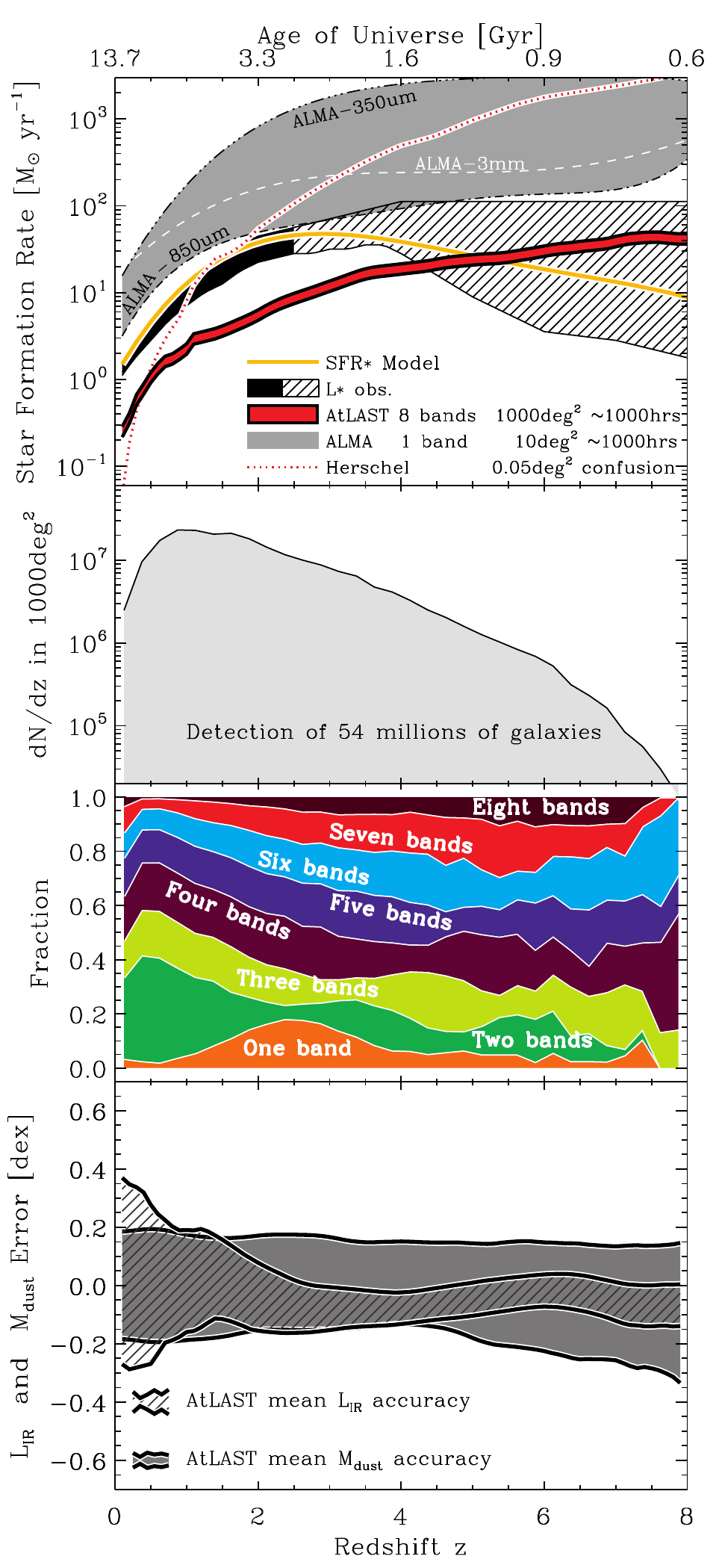}
    \caption{Exploring the parameter space of a 1000 deg$^2$ survey using 1000 hours of AtLAST observing time. Top panel: SFR vs. redshift survey limit. Second panel: $dn/dz$. Third panel: fraction of (mock) sources detected when multiple bands are used. Bottom panel: expected AtLAST accuracy in inferring $L_{FIR}$ and $M_{dust}$ as a function of redshift, demonstrating why multi-band detections are important.
    \citep[Figure reproduced from ][]{vanKampen2024}}
    \label{fig:survey1000}
\end{wrapfigure}

\subsection{Distant Universe and Cosmology}
\label{SciTheme_Distant}

\subsubsection{Unbiased Surveys of the Distant Universe}
\label{SciTheme_DistantSurveys}
\textit{See \citet{vanKampen2024} for further details.}

Single dish sub-mm telescopes like the JCMT were the first to identify an entirely new family of galaxies in the distant universe: dusty star forming galaxies (DSFGs, also known as sub-mm galaxies), which are too deeply embedded in their own dust and gas to be detectable at shorter wavelengths \citep{Blain2002, Casey2014}. They are the most actively star forming galaxies in the early universe, responsible for the bulk of star formation over cosmic time, and seem to be more abundant than expected from models. Much of our understanding of them as a population has come from sub-mm observations - whether through the pioneering sub-mm single dish and interferometric observations from the ground (e.g. JCMT, CSO, SMA, ALMA), or from space (i.e. Herschel, Planck). These surveys are currently biased towards the most massive or gas-rich galaxies \citep[10-100x more massive than the Milky Way;][]{Hodge2020}. Because of the confusion limits of current facilities, we cannot currently probe the underlying population of `normal' galaxies, i.e. the galaxies that constitute the bulk of the sources contributing to the cosmic infrared background (CIB) in an unbiased manner. Understanding this population is important for understanding the relationship between star formation and black-hole growth \citep{Carraro2020, Mountrichas2023} and the evolution of dust over cosmic time \citep{Sommovigo2022, Dayal2022, DrewCasey2022, Hirashita2022, DiCesare2023}. 


In order to create blind, unbiased spectroscopic surveys of the ‘normal’ population of dusty galaxies at high redshift requires a telescope with a high throughput, and the sensitivity and resolution that come with a large primary mirror.  With highly multiplexed instrumentation on AtLAST, this will, for the first time, be possible.  With the types of surveys described below, we will be able to derive number counts, clustering estimates down to small scales and redshift distributions of the normal population of dusty galaxies down to $L^\star$, the luminosity of a ‘typical’ galaxy\footnote{More precisely, it is the characteristic luminosity at which the luminosity function exhibits a rapid change in slope \citep{Schechter1976} and it happens to be very close to the luminosity of the Milky Way.}. Such a survey, functionally similar to the SDSS \citep{Geach2019}, would complement upcoming optical and near-IR surveys like those from the Rubin Observatory 
as it would allow us to measure redshifts as well as the gas and dust contents of hundreds of thousands of galaxies, many of which will have optical counterparts detected by the Rubin Observatory.


As confusion limits are proportional to the resolution of the telescope, AtLAST's 50~m aperture means its confusion noise limits at sub-mm wavelengths will be much better (i.e. lower noise) than any other current or planned observatory. For instance, the photometric confusion noise at 350~$\mu$m (see Appendix \ref{app:confusion}) will be a factor of over 10,000 lower than that of 6~m telescopes like ACT and FYST. With a multi-chroic camera, a 1000~deg$^2$ continuum survey could be conducted in 1000 hours that reaches a sensitivity limit of 570~$mu$Jy at 350~$\mu$m, thereby resolving 82\% of the CIB. This will give the most comprehensive view of the infrared luminosities and dust masses of galaxies, measure star formation rates out to a redshift of $z$=5, and give new insights into galaxies at a wide range of redshifts, masses, metallicities, environments and morphologies (see Figure \ref{fig:survey1000}). With the appropriate cadence, this will also enable transient science (see Section \ref{SciTheme_Transients}).

To accurately measure the redshifts of these galaxies requires detecting spectral lines, which requires much deeper observations.
Taking a ‘wedding cake’ approach enables smaller area, deep spectroscopic studies within regions of the larger continuum survey described above. By layering from shallower but larger area surveys to deeper and smaller areas we build up a deeper understanding of an unbiased sub-sample of the normal dusty galaxy population, pushing out to redshifts as high as 7 in these deeper spectroscopic layers. In a 1~deg$^2$ area, it is expected that around 100,000 galaxies would be detected in around 3000 hours. In addition to measuring their redshifts, this would provide information on the gas content, cooling budget, star formation rate, dust mass and dust temperature of these galaxies.

Methods like line intensity mapping (LIM, to probe the power spectra of galaxies), Baryon Acoustic Oscillations \citep[BAO, for measuring fluctuations in the density of baryonic matter in the primordial universe][]{Cole2005} and related redshift space distortions \citep[RSD][]{Peacock2001} using low resolution spectrometers  can also be derived from these surveys to address the tension between the Hubble constant measured at different cosmic epochs \citep{DiValentino2021}. With a spectroscopic survey of [CII] and CO covering 1000~deg$^2$ it will be possible to measure the Hubble constant to within 0.7\% and the growth rate  to a precision of 7.3\%.

Galaxy evolution in proto-clusters is currently not well understood due to the small and heterogeneous samples from current surveys. A distinct large survey of distant galaxy clusters with AtLAST will produce a high-redshift counterpart to local large surveys of rich clusters like the well-studied Abell catalogue, to understand the impact of environment on the formation and evolution of these distant cluster galaxies. To guarantee spectroscopic redshifts from $z=4.5-10$ and a comprehensive study of the most prominent lines emitted from the cold ISM, an instrument with spectral coverage from 180 to 345~GHz will be necessary. If this is extended up to 700~GHz then it will be possible to map the peak of the star formation and black hole activity of the Universe at $z=2$ \citep{Madau2014}.

\subsubsection{The Warm and Hot Universe via the SZ Effect}
\label{SciTheme_SZ}
\textit{See \citet{DiMascolo2024} for further details.}

Warm and hot ($\gtrsim10^5$~K) ionised gas makes up a large part of the matter budget across a range of scales, from intergalactic filaments, to the intracluster medium (ICM) and down to the circumgalactic medium (CGM) of individual galaxies. This hot gas provides a powerful way of understanding the interconnected evolution of populations of galaxies and large-scale cosmological structures, the matter assembly of the Universe and its thermal history. Such hot gas has typically been the realm of x-ray observations, but the Sunyaev-Zeldovich (SZ) effect provides a way for this gas to be studied in the \submm{}. This is a redshift independent spectral distortion of the cosmic microwave background (CMB) through inverse Compton scattering by high energy electrons \citep{Sunyaev1970, Sunyaev1972, Sunyaev1980}, with the most dominant contributions coming from its thermal and kinetic components (referred to as tSZ and kSZ, respectively). 

The tSZ effect is the distortion of the CMB spectrum due to thermalized reservoirs of high-energy electrons -- as, e.g., in galaxy clusters --, resulting in a shift in intensity at higher frequencies and is sensitive to ICM pressure. The kSZ effect is the Doppler shift in CMB photons caused by the proper motion of hot electron halos (or part thereof) relative to the CMB rest frame and is sensitive to their peculiar velocity.
These effects 
distort observed emission over wide frequency ranges, which requires the use of highly sensitive multi-chroic cameras that operate across the (sub-)mm wavelength range to be detected. The targets of interest -- from individual galaxies and galaxy clusters to large-scale cosmic web -- are diffuse and extend across many degrees, which requires large fields of view and highly sensitive observations. A telescope like AtLAST, with its large aperture, large FoV and instrumentation across the (sub-)mm wavelength range would be ideal for using the SZ effects to probe fundamental physics like the thermodynamical state of the ICM, the warm/hot component of the circumgalactic medium (CGM) and warm/hot intergalactic media, when they form, how they evolve and what impact they have on star formation and galaxy evolution \citep[see e.g.][]{Mroczkowski2019}, as described below.

Observations of the morphology and thermodynamics of the hot gas in galaxies, groups, and clusters gives a historical record of the physics that shaped it (i.e. AGN feedback, dynamics, mergers, accretion and cooling). The effects of these physical processes can be seen as deviations from radially symmetric thermodynamic structure (predicted by simple models of gravitational collapse) in the pressure of the hot gas. The tSZ gives a direct proxy for the pressure deviations and measuring the tSZ effect gives us a calorimetric view of the thermal properties of the hot gas hosted within haloes of individual cluster galaxies, the extended ICM, and out to the large-scale IGM outskirts. 
Higher sensitivity and greater resolution with AtLAST is required to get to the point of being able to detect the full extent of the pressure distributions in the ICM of galaxy clusters. With that information, we will probe how the pressure distributions in high-z galaxy clusters deviate from universal pressure models (which feeds into understanding multi-scale processes like mergers) and how the pressure profiles in the outskirts of clusters vary, which feeds our understanding of how well-virialised these clusters are, and where their accretion shocks are. Understanding these processes is key to our understanding of large-scale structure formation in the Universe.

On smaller scales, a calorimetric view of the gas in a galaxy tells us about feedback processes on the size scales of galaxy haloes to full clusters. For instance, AGN-driven outflows contribute to heating on these scales, and it is still not clear how and through what specific mechanisms supermassive black holes contribute to evolution of their host galaxies. Current studies of `integrated' tSZ effects are starting to probe the imprints of these feedback mechanisms \citep{Crichton2016, Hall2019,Yang2022}, but are hampered by the necessity of stacking the data, which dilutes their usefulness.

Including the corrections for relativistic effects on the tSZ effect, we can directly measure the temperature of ICM electrons. Adding this temperature measurement to the pressure and density measurements derived above, offers the ability to measure entropy in the ICM, which is modified by the evolutionary effects / physical processes mentioned above. This significantly improves our physical models of galaxy clusters and their use as cosmological probes. Measuring the relativistic thermal SZ effect for individual targets is generally beyond the capabilities of current facilities. Higher resolution, highly sensitive observations across the (sub-)mm band are required to reach the sensitivities necessary to detect this effect. 



Using instead the kinematic SZ (kSZ) effect gives insights into the peculiar motions of cosmic structures like galaxy clusters. However, the shape and the relatively weaker signal in hot clusters can make it more challenging to detect, as does the fact that the spectral distortion characteristic of the kSZ effect is consistent with Doppler shifting of CMB photons, which makes it hard to distinguish from CMB anisotropies. In order to break this degeneracy in individual observations, sensitivity to a wider range of spatial scales (300 $\lesssim \ell \lesssim$ 20000 in terms of spherical harmonics) than possible with current or planned facilities (including dedicated CMB facilities) is required.\footnote{Note, however, that pairwise stacking has been used extensively to separate the kSZ from both CMB and tSZ signals in a statistical sense.} Further, such observations benefit from a 220 GHz channel, where the kSZ signature is dominant. Once detectable, the kSZ effect can be used to probe the amplitude and growth rates of cosmological density perturbations which then translates to being able to distinguish $\Lambda$CDM from alternative cosmologies \citep{Kosowsky2009, Mueller2015, Bianchini2016}. Further to this, detailed spatial mapping can be used to characterise turbulent motions and thus the important driving dissipation scales relevant for the feedback processes.

Being able to study the warm-hot intergalactic medium (WHIM) in detail through both the tSZ and kSZ effects will give access to a better understanding of the `missing baryon' budget as the dominant fraction at $z<3$ is expected to be beyond the virial radii of their host galaxies, where the WHIM is (i.e. in the Cosmic Web). To date, the most accurate way of probing the (high temperature end of the) WHIM is through stacking of SZ images and X-ray measurements. The latter suffers from the inherent biases described above, while the former suffers from averaging out details in the process of stacking. 

\subsubsection{The Hidden Circumgalactic Medium}
\label{SciTheme_CGM}
\textit{See \citet{Lee2024} for further details.}

Galaxies are surrounded by a reservoir of gas and dust extending beyond the interstellar medium (ISM) out to the virial radius, commonly referred to as the circumgalactic medium (CGM\footnote{We do not impose a strict definition of the CGM. In our loose definition, the CGM includes the matter at the boundary between ISM and CGM.}). Understanding the CGM allows us to evaluate the feedback and feeding mechanisms that impact the galaxy's ability to sustain star formation. These feedback mechanisms leave observable signatures in the density, temperature, metallicity, and morphology of the CGM. The total mass of the CGM reservoir and the contribution from different gas phases (neutral/cold versus highly ionised/hot) that exist within it are largely unconstrained, as are the physical processes that affect and shape the CGM. Observing the CGM is challenging because of its primarily faint, diffuse, and extended nature (the emission can extend to many hundreds of kiloparsecs -- corresponding to an angular size of degrees for galaxies within 10~Mpc of the Milky Way), with the ability to resolve sub-kiloparsec scale clumps and a dynamic range capable of distinguishing the CGM from the significantly brighter ISM.


As suggested above, there are a number of key areas where probing the CGM can better constrain models of galaxy formation and evolution. The first is the nature of cosmic accretion, which digs into how galaxies can sustain high levels of star formation across cosmic time. It is still unclear how fresh material is accreted onto a galaxy to sustain this star formation; is it through direct accretion of cosmic streams in the intergalactic medium (IGM), or is it reprocessed through the CGM? The second is the missing baryon problem, which focuses on reconciling the difference between the expected baryon mass of galaxies and their measured values. Is this missing baryonic mass, which is comparable to or can even excess the amount in the disc itself, hidden in the cold CGM? The third is the impact of feedback on galaxy evolution.  We know that feedback from star formation and AGN must play a role in galaxy evolution, but what is their cumulative effect over time and size scales? The answers to all these questions are likely hidden in the CGM.

Studies of the CGM in distant (z$\geq$1) galaxies have discovered large reservoirs of cold gas surrounding the discs of galaxies, especially around quasars and dusty star-forming galaxies (DSFGs, see Section \ref{SciTheme_DistantSurveys}). Protoclusters at higher redshift (z$\sim$4) show starburst events coordinated across galaxies separated by hundreds of kiloparsecs, hinting at the large scale gas supply on to the protocluster region to sustain the massive starburst configuration. 

In clusters, Brightest Cluster Galaxies (BCGs) and interacting galaxies at intermediate to high redshift seem to show these types of gas reservoirs \citep[e.g.][]{McDonald+14,Dunne+21}, sometimes extending to tens of kpcs, which is at odds with the relatively gas-poor galaxies found in dense clusters in the local universe \mbox{\citep[e.g.][]{Haynes1984,Zabel2019}}. In the local universe, galaxy interactions, tidal tails and streamers are indicated as extended gas reservoirs and they could supply gas to feed star formation and AGNs. Further, the feedback from star formation and AGNs through galactic outflows and heating changes the chemical and dynamical properties of the CGM (over time). However, observing the CGM is tricky because of its highly extended nature and inherently low surface brightness.

\begin{figure*}[tbh] 
\centering
   \includegraphics[width=0.9\textwidth]{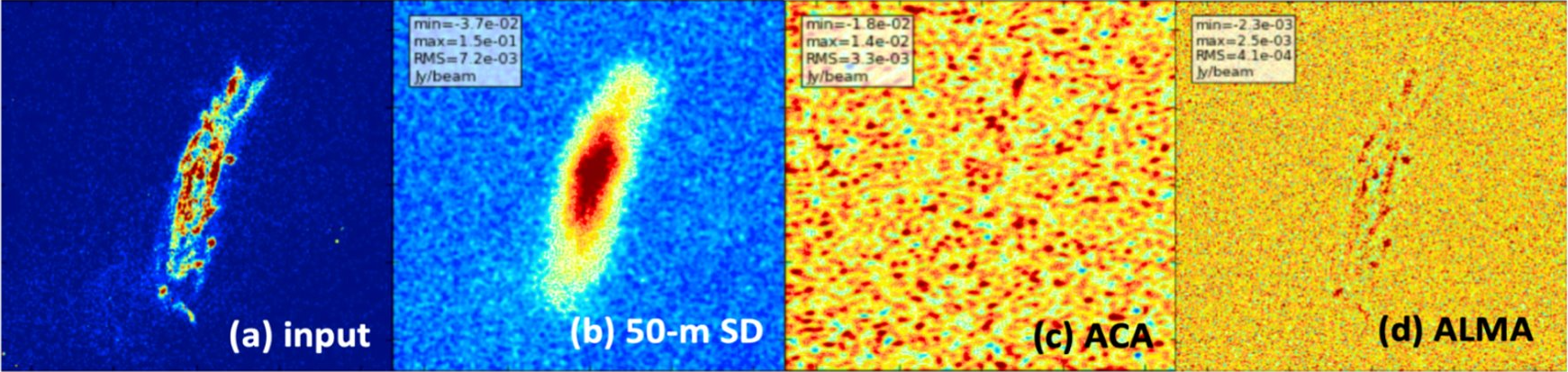}
      \caption{Mock observations of CO(3-2) emission from the ISM and CGM of a star-forming galaxy at $z=0.02$ with an extent of $\sim$80~kpc (200''). The input simulation is shown in panel (a). Panels (b), (c) and (d) show observations assuming 10 hours of on-source time for a 50~m single-dish using a single pixel ALMA detector, ACA and ALMA respectively. \citep[Figure reproduced from][]{Cicone2019}}
   \label{fig:CGMsim}
\end{figure*}

 Observations with current facilities, such as JCMT, APEX, ALMA and ATCA have shown hints of extended atomic and molecular gas emission around galaxies across cosmic time. However, their low surface brightness and their large emitting areas (especially for nearby galaxies) make detailed observations challenging, if not impossible with current facilities (see Figure \ref{fig:CGMsim}). For interferometers, the large scales involved in CGM emission prove difficult to detect 
 and current generation single dish facilities do not have the sensitivity required to probe this emission. These two aspects combine to show the need for a new facility, a 50m class single dish facility with a large field of view like AtLAST, with the sensitivity to detect low surface brightness emission on size scales of up to a degree. Current facilities are able to pick up some of the brightest components of the CGM, such as the tidal streams and fountains mentioned above, but cannot yet detect the underlying gas mass reservoir, the extent and detailed chemical composition.

High-frequency forbidden lines (i.e. [CI], [CII] and [OIII]) are good tracers of the CGM of distant galaxies \citep{Schimek2024b, Schimek2024} and [CI] is still detectable in the sub-mm even for nearby galaxies, although requires a very high sensitivity that is only possible with a large aperture. 
This motivates the need for the ability to detect this line at low redshifts, 
which in turn sets a requirement for the telescope to be able to observe in the highest frequency sub-mm atmospheric windows. In addition to these spectroscopic observations, observing the continuum emission from dust across the (sub-)mm wavelength regime probes the underlying dust content of the CGM as well. 

\subsection{Nearby Galaxies}
\label{SciTheme_NearbyGal}
\textit{See \citet{Liu2024} for further details.}

Studying the physics and chemistry of nearby galaxies allows us to understand the interplay between the many processes that govern galaxy evolution across a much wider range of physical environments and extremes than in our own Galaxy, but at resolutions as close to Galactic as possible. Star formation theories based on Galactic observations cannot convey the complexity or diversity of environments in which other galaxies form and evolve (e.g. with different metallicities, total masses, positions within and size of galaxy cluster, etc.).  

The closest companions to the Milky Way, the Magellanic Clouds (MCs), offer a unique perspective on low-metallicity environments with near galactic scale resolutions.  From current generation facilities, we can quantify their physics and chemistry on sub-parsec scales and understand, in general terms, the relative contributions of the different components of their ISMs. But observations of [CII], for instance, suggest much of the molecular gas in the Large Magellanic Cloud (LMC) is CO-dark: we are not detecting it using standard sub-mm methods \citep{Chevance2020}.  Understanding the molecular gas content of the LMC then leads us to understand how much star formation can be expected, and whether that is driven by the contents of the observed giant molecular clouds, or the ISM surrounding them. Current facilities do not have the sensitivity, FoV and resolution to explore the CO-dark gas in the MCs. With the ability to observe in the sub-mm, AtLAST will be capable of observing [CI] lines (at rest-frame frequencies of 492 and 809~GHz) across the MCs that trace this CO-dark gas.

\begin{figure*}[hbt]
    \centering
    \includegraphics[width=0.47\textwidth]{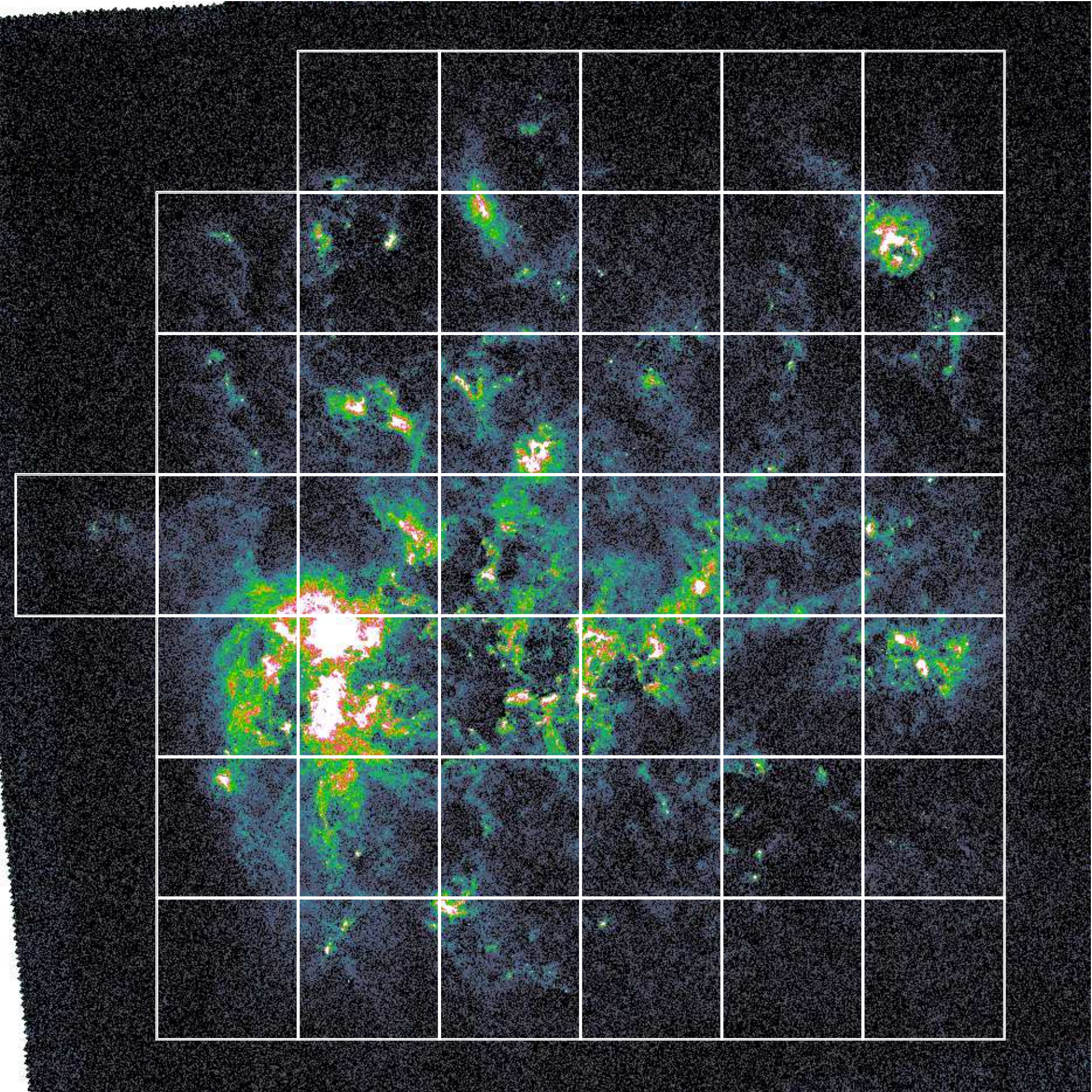}
    \includegraphics[width=0.47\textwidth]{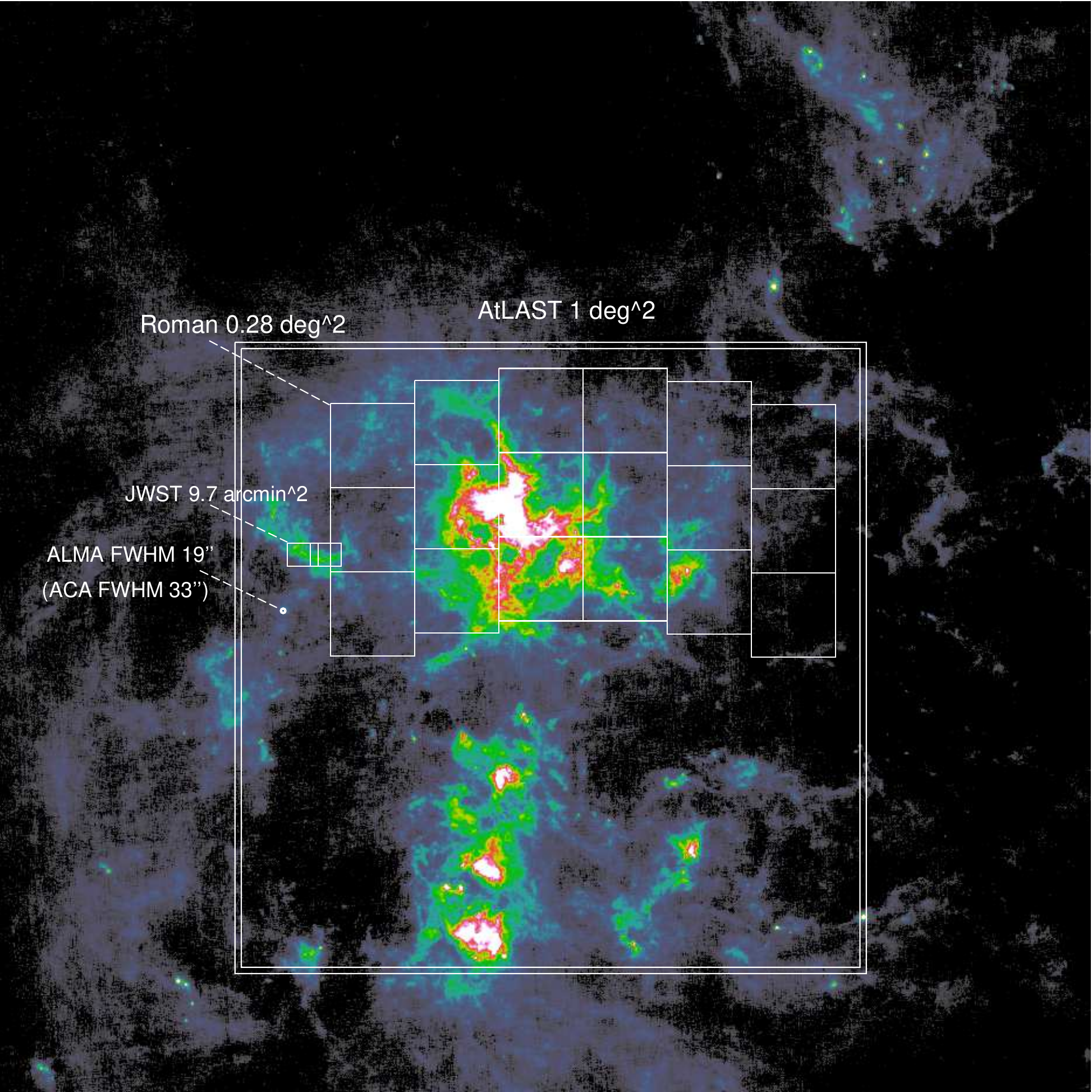}
    \caption{Dust emission at 100~$\mu$m in the LMC from the \textit{Herschel} space observatory \citep{Meixner2013} The left panel shows the full $\sim 7^{\circ} \times 7^{\circ}$ area of LMC. The grid boxes represent the $1^{\circ} \times 1^{\circ}$ fields of view proposed for the AtLAST continuum camera. The right panel shows the zoom-in $30'$ view around the 30 Doradus H{\textsc{ii}} region. The fields of view of AtLAST, ALMA (12-m array and 7-m array), \textit{JWST} NIRCam and \textit{Roman} Space Telescope WFI are shown for comparison. Only the future-generation AtLAST and \textit{Roman} are able to map the full near to far-infrared and sub-millimeter dust emission in the LMC. \citep[Figure reproduced from][]{Liu2024}}
    \label{fig:lmc-herschel}
\end{figure*}

Quantifying the dust (in terms of mass, density and temperature) as a function of the environment, energy injection and metallicity tells us about dust properties in extreme environments. There are discrepancies between theoretical expectations of dust properties at low metallicity and those seen in the MCs \citep{Bot2010,Galliano2011}. Resolving this discrepancy pushes us closer towards a complete model of cooling, heating, star formation and dust grain growth in different metallicity environments. Understanding low-metallicity environments in nearby galaxies allows us to improve our understanding of high-redshift galaxies, which are typically also lower metallicities.

Pushing out to larger distances, the physics and chemistry of the ISM in nearby galaxies tells us about the evolution of that galaxy, and by deriving both the star formation rate and gas surface density in nearby galaxies \citep[the Kennicutt-Schmidt relation;][]{Schmidt1959,Kennicutt1989}, we understand different types of galaxies in the context of the general population of galaxies.  These types of relations break down on roughly kpc scales, and understanding that deviation from the KS relation puts interesting constraints on the timescales of star formation.  Through observations of chemical tracers tracing different densities and in multiple transitions, we can further explore the parameter space in the KS relation, probing to higher and higher gas densities.  To do this well, in multiple chemical tracers, across statistically significant samples of galaxies, requires highly multiplexed spectroscopic instruments with high sensitivity and resolution.  

Threading through these themes is the ability to measure the polarised emission of these galaxies to quantify how magnetic fields influence the evolution of galaxies.  Magnetic fields regulate star formation, influence the dynamics of the ISM and are important for a number of other processes, but observing them consistently remains a significant challenge.  Small samples of polarimetric observations of dust in nearby galaxies have shown a relationships between the fields in the molecular gas and star forming regions, but measurements are difficult because of either the small fields of view of sensitive interferometers like ALMA, or the relative insensitivity / low resolution of current generation single-dish facilities where observing polarisation fractions to uncertainties of a few percent requires a significant investment of telescope time.

Creating a polarisation survey of 100 nearby galaxies, reaching resolutions of 20-300 pc to an rms noise level of 16 $\mu$Jy/beam will enable detections of 1\% polarisation in the dust emission. This is 100 times deeper than possible with current facilities while at the same time offering significantly higher spatial resolution (by a factor of $\sim$3 in each direction) which means more than 1000 independent polarisation vectors can be measured. While impossible with current facilities, observations like this could be done with AtLAST in 18-50 hrs per galaxy.  

\subsection{Our Galaxy}
\label{SciTheme_MilkyWay}
\textit{See \citet{Klaassen2024} for further details.}

From the vantage point of our Solar System being in the plane of the Milky Way, we are able to probe the cycle of interstellar matter
in a detailed way that is impossible to capture for external galaxies. We can view the assembly of clouds on the largest scales down to dense cloud core scales and protostars, then follow the subsequent dispersal of clouds by the feedback of the formed stars and the eventual enrichment from the death throes of stars at the end of their lives. We can delve into the minutiae, but to put those details into their galactic context requires large area surveys. With a large aperture, wide-field of view (sub)millimeter single dish telescope, we can explore and quantify the spatial distribution of gas and dust, the kinematics of the gas, the gas chemical composition (and its evolution) and magnetic fields from the scales of giant molecular clouds down to individual nearby, faint debris discs, analogous to our own Kuiper belt.

Large scale surveys of the gas and dust in the Galactic Plane unlock a multitude of science questions.  How do cool Giant Molecular Clouds (GMCs) form within the largest scale structures in the ISM of our galaxy? How do those GMCs then continue to fragment into filaments and star forming clumps and cores? What then triggers or inhibits star formation in those clumps and cores? What is the interplay between magnetic fields and stellar feedback on star-formation? From these questions, we can see that the chemistry, dynamics, and recycling of material are all interrelated in our understanding of the ecology of our Galaxy and tie directly to the gas, dust and magnetic field properties of the ISM. The ISM is turbulent and magnetized and to describe the evolution of the ISM and star and planet formation we need to probe the density distribution of matter as well as the velocity structure (to probe the dynamics) and the magnetic fields. AtLAST will be able to study simultaneously the evolution of dust and gas from large scales to small  (from $\sim$100 pc to $<$0.1~pc) and from low densities to high (1 to 10$^6$ cm$^{-3}$) spanning orders of magnitude in scale and densities.

A polarised dust continuum survey of the Galactic plane will allow us to constrain, in a statistical and holistic way, the relationship between gravity and magnetic fields in the collapse and support of structures, from the scales of entire GMCs down to those of star-forming cores. This will allow us to understand the flow of material through and between states, and the nature of star forming cores throughout the plane -- as we will reach from filament to individual core scales (0.1 pc) resolution on the far side of the Galaxy at the highest frequencies obtainable with AtLAST. 

Spectroscopic surveys of the Galactic plane add to the picture through chemistry, turbulent velocity dispersions, temperatures and heating and cooling rates of the gas. Wide bandwidths will be essential for observing multiple species and transitions simultaneously.  Here we can use a similar `wedding cake' approach to the Distant Universe studies above. In a wedding cake scenario, we begin with larger area surveys (i.e the whole Galactic Plane) to sensitivities sufficient to detect common molecular and ionised gas transitions (CO, CS, N$_2$H$^+$, HCO+, SiO, radio recombination lines, etc), and then follow up with deeper / more sensitive observations in a smaller tier towards  regions deemed chemically complex (i.e. individual star forming regions) in order to derive a census of complex organic molecules (defined as carbon bearing species with at least six atoms). Only following up with highly sensitive observations in regions known to be chemically complex will save hundreds of hours of observing time.

With the resolution, sensitivity to the full range of spatial scales, and confusion limits enabled by a 50m dish antenna, we can study the most common mode of star formation: low mass stars in clustered environments which are also in the process of forming high mass ones.  With current technologies, we can study the details of isolated core collapse in nearby low mass star forming regions, but with a 50m class telescope, the same resolution achieved in nearby (100-200 pc) regions will be possible in the richer and more complicated high-mass star forming regions within 2 kpc - where a full Initial Mass functions (IMF) worth of low and intermediate mass stars are also forming but cannot currently be detected or disentangled from their brighter companions.

\begin{figure*}
    \centering
    \includegraphics[width=0.9\textwidth]{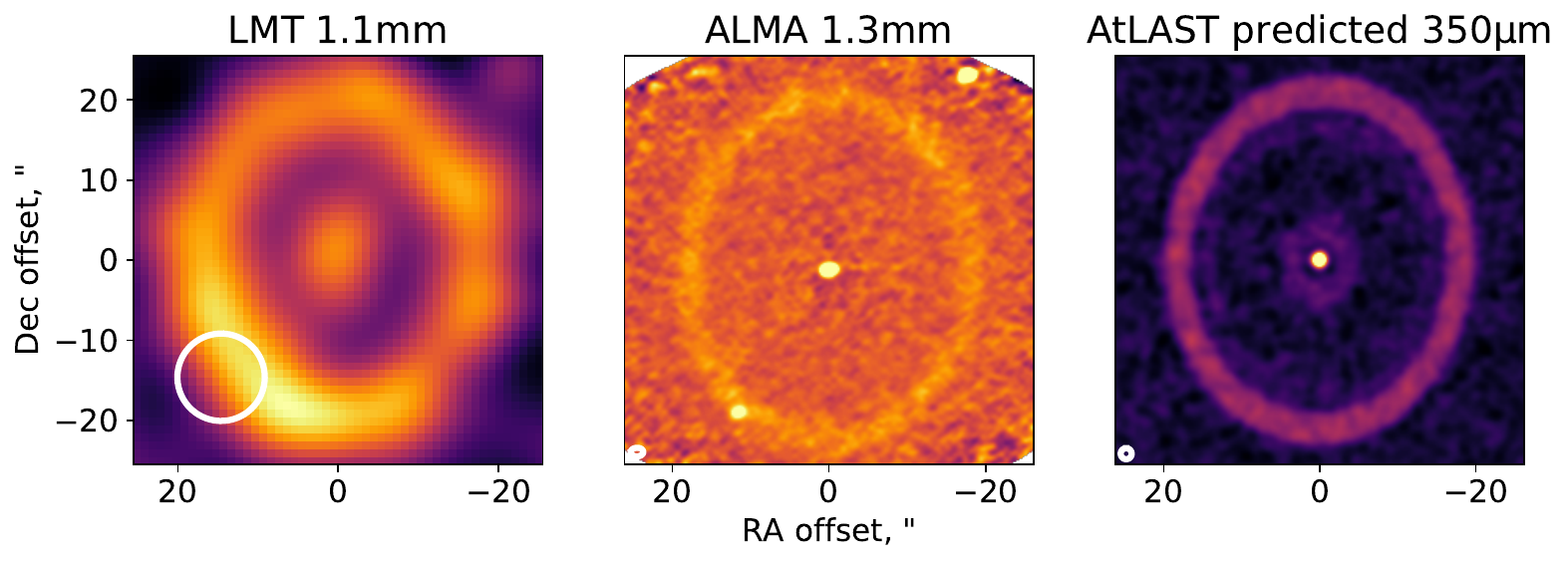}
    \caption{Observations of $\epsilon$ Eridani with LMT \citep{2016Chavez} and ALMA \citep{2023Booth} along with a predicted image using AtLAST for observation with an on source time equivalent to the ALMA image. For a nearby disc like this, AtLAST can reach scales of a few au much more rapidly and without loss of large-scale flux that hinders such observations with ALMA. \citep[Figure reproduced from][]{Klaassen2024}
    }
    \label{fig:epsEri}
\end{figure*}

With large format cameras, entire nearby star forming regions can captured in a single pointing, taking a single `snap-shot'. This makes it possible to not only trace the dust and gas in the star forming cores that then settles into protoplanetary discs and as the stars emerge and finally evolve into debris discs as the gas is dispersed from the system, but to go back to these regions in a set cadence to monitor for transient behaviour (i.e. accretion flares). With \submm{} observations of the early evolution of these planetary systems, the masses, morphologies, grain-size distributions and dispersal mechanisms can be tracked \citep{Lovell2021c}. With repeated observations of these star-forming regions we can also expect to find large numbers of variable sources, informing us of the star formation process (see also Section \ref{SciTheme_Transients}). The mass assembly of protostars is known to be episodic and for the youngest, most deeply embedded sources, during the main mass assembly stage, brightness variability in the far-IR and sub-mm provide the only direct measure of both the timescale and amplitude of these accretion episodes - only AtLAST has the ability to monitor 1000s of sources with monthly cadence over decade timescales.

As the circumstellar discs evolve into debris discs, their dust masses greatly decrease, meaning that only the brightest debris discs are detected \citep[e.g.][]{Wyatt2008}. Observing discs as faint as the Solar System's Kuiper belt, requires observing the nearest stars (within around 20~pc) down to sensitivities on the order of 10s of $\mu$Jy/beam. Given their proximity, the discs can easily extend well beyond the maximum resolvable scale of interferometers like ALMA and so only a large single-dish telescope is capable of detecting them (see Figure \ref{fig:epsEri} for an example). Observations in multiple bands then allow the characterisation of their dust properties \citep[e.g.][]{2020Lohne}.

\subsection{Solar System}
\label{SciTheme_Solar}
\textit{See \citet{Cordiner2024} for further details.}

Understanding our place in the Universe requires understanding the formation and evolution of our own Solar System. (Sub-)mm spectroscopy of giant planets teaches us about the chemistry and dynamics of their atmospheres, from which we obtain information about their origins and temporal evolution \citep{encrenaz2005}. From spectroscopy of terrestrial planets, moons and comets, we can learn about their habitability and the availability (and origins) of chemical ingredients for life in the Solar System. Deeper insights into our own planetary system are also essential in our quest to understand the properties of exoplanets, and the relationship of our solar system to other planetary systems.

\begin{wrapfigure}[24]{r}{0.5\textwidth}
\centering
\includegraphics[width=0.48\textwidth]{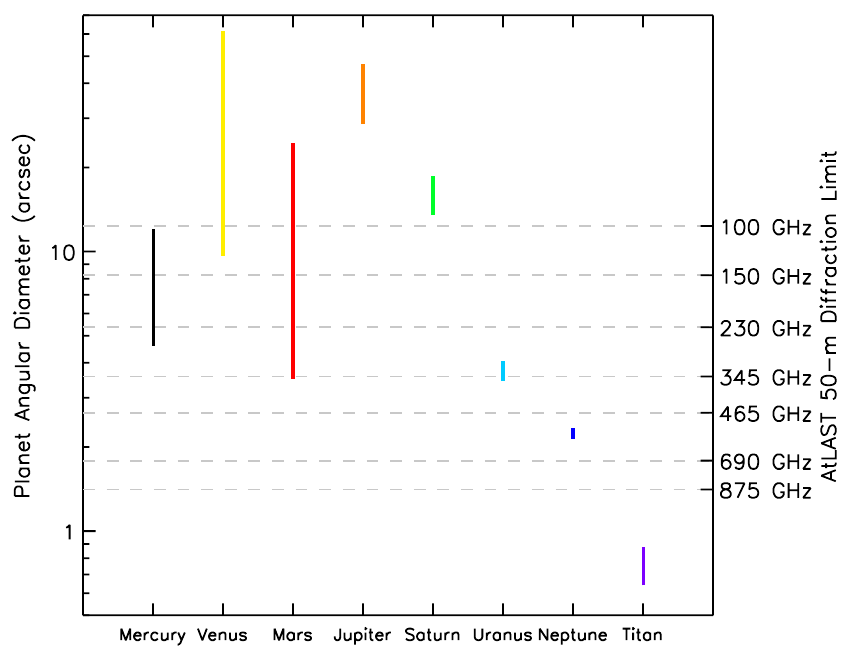}
\caption{\label{fig:res} Angular diameters of major Solar System bodies and Saturn's largest moon, Titan. The vertical extents of the coloured bars for each body represent their range of angular sizes due to differing geocentric distances throughout the year. The second y-axis shows the angular resolution a 50 m single dish facility would achieve for a range of representative frequencies. \citep[Figure reproduced from][]{Cordiner2024}}
\end{wrapfigure}

Observations with current (sub-)mm facilities have produced ground-breaking results, including measuring the molecular inventories and dynamical motions of planetary and cometary atmospheres. However, current observatories suffer from limited dynamic range, spatial and temporal coverage, and sensitivity \citep{Cordiner2024}. Instantaneous mapping over the entire face of a body is critical for detailed studies of the rapidly rotating and evolving atmospheres of solar system bodies \citep{lel19,cord20}.

Interferometric observations are not ideally suited to solar system studies due to the spatial filtering of the large angular scales that are a common feature of planets and comets. Furthermore, the repositioning of antennas (and consequent variation in $uv$ coverage) hinders our ability to reliably and consistently study and map temporally variable phenomena such as storms, outbursts or seasonal changes.

In order to advance our knowledge of planetary atmospheres, a large aperture, single-dish telescope is necessary. This will allow us to measure spatial and temporal variations in temperature structure and molecular abundances throughout the troposphere and stratosphere, which can be used to investigate the circulation, dynamics and composition of the atmosphere. A sensitive single-dish facility will enable short-term and seasonal variations to be studied, and facilitate the discovery of new, trace molecular species and their distribution. A 50~m dish is essential here so that Venus, Mars, Jupiter and Saturn can be resolved throughout their orbits and across all wavelengths (see Figure \ref{fig:res}). With this, Uranus and Neptune can also be marginally resolved at the shortest wavelengths. In order to conduct sensitive searches for new spectral lines and isotopologues of interest, a spectroscopic dynamic range of $\sim10^5$ is needed due to the continuum brightness of the planets --- to realise this, close attention should be paid to achieving the flattest possible spectral bandpass through the careful choice of optical design, receiver components and optimal calibration strategies.

Moons of the giant planets also offer compelling targets in the search for life elsewhere in the solar system. Titan has a thick, carbon and nitrogen-rich atmosphere that can be studied in a similar way to the terrestrial planets \citep{Nixon2024}. Deep spectral line observations will enable the discovery of new organic molecules \citep{cor15,pal17,the20}, 
as well as isotopologues of known molecules that improve our understanding of the long-term physico-chemical evolution of Titan's atmosphere \citep{man09,man14,nom23}. 
Many of the icy moons of the outer planets are thought to have subsurface oceans, which can result in cryovolcanism \citep{nim16}. Spectral studies of the resulting plumes provide the perfect opportunity to study the compositions of the subsurface oceans, in particular searching for organic molecules that can reveal insights into the habitability of these oceans. The Cassini mission to Saturn discovered such plumes around Enceladus \citep{Hansen2006,por06,spe18}. 
Although it had a mass spectrometer, this lacked the resolution to differentiate between certain molecules \citep{wai06, wai09}. 
Remote observations from ground-based facilities present the opportunity to overcome this. However, to do so requires a high enough angular resolution (specifically $<8''$ at 1.1~mm) to avoid interference from Saturn and its rings, and with a single-dish to avoid any interferometric artifacts. This would enable both confirmation of the Cassini result and continued monitoring for temporal changes in the plume composition.

Comets are thought to be largely unchanged since they accreted very early in the Solar System's history \citep{mum11} 
and thus provide unique information on the physical and chemical conditions at the time of planet formation \citep{boc04}. The ices and organics contained within comets may well have played a key role in setting the stage for life on Earth due to their delivery via impacts. Studying comets is also useful for understanding how our Solar System relates to other planetary systems and to the chemistry of the ISM (see Section \ref{SciTheme_MilkyWay}). A huge variety of different ice and organic lines are detectable in the (sub-)mm, necessitating a 32-64 GHz bandwidth to be able to observe them simultaneously and thus avoiding changing tunings during time-critical observations. Of particular interest is the deuterium to hydrogen ratio that varies across Solar System bodies and can tell us about the history of water in the Solar System, particularly when compared to observations of protoplanetary discs \citep{wil09,alb14,cle14}. 
The strongest HDO line observable from ground-based (sub-)mm observatories is the 894~GHz line, demonstrating the importance of reaching these high frequencies. Cometary comae are often extended over a few arcminutes, making studies of the extended coma impossible with interferometry. In order to resolve the complex kinematical structure of the coma, a spectral resolution of $\sim0.1\,\rm km\,s^{-1}$ is necessary.

\subsection{The Sun}
\label{SciTheme_Sun}
\textit{See \citet{Wedemeyer2024} for further details.}

The Sun is the only main sequence star we can spatially resolve, and as such, not only does it allow us to understand the engine that drives our solar system, but gives us insights into other stars that would be impossible to observe in such detail.  At (sub-)mm wavelengths, the emission from the Sun is dominated by the magnetised plasma in the chromosphere, which we observe as (polarised) thermal continuum emission. Different layers of the chromosphere dominate at different wavelengths, hence observations across the (sub-)mm regime probe the three-dimensional structure of the Sun's atmosphere. Observations of the chromosphere are important for a number of reasons, including understanding the `coronal heating problem' -- understanding why the corona is much hotter than the solar atmospheric layers below it (including the photosphere and chromosphere), solar flares, coronal mass ejections (CMEs), prominences and the solar wind. By disentangling the importance of different chromopsheric processes (such as magnetic reconnection and various forms of wave heating), we start to unpick which types of physical processes are dominant under which conditions (i.e. active or quiet Sun) and on what size scales.

Solar flares, prominences and the solar wind all play a vital role in space weather.  Predicting them and understanding what drives them will help shape space weather forecasting, but doing that requires detailed long term observations of the Sun in polarised light.  Because these transient phenomena are related to the magnetic field reconnecting and reconfiguring on sometimes explosive timescales, simultaneous observing of the full face of the Sun is required, with repeated observations to take in variations across the solar rotation period (around a month) and solar cycle (11 years).

Space weather directly impacts human society due to the effect it has on power grids, satellite infrastructure and human spaceflight. The more we understand space weather, the better we can mitigate these effects. Understanding space weather in our own Solar System also allows us to better understand space weather in extrasolar planetary systems and how it impacts habitability in those systems, particularly in cases where flares and CMEs are considerably stronger than those released from the Sun \citep[e.g.][]{Maehara2012}.

Current generation facilities often either do not have the spatial resolution to resolve structures like granules \citep{Kundu1959,Trottet2011, Loukitcheva2014} or, as is the case with ALMA and other interferometers, filter out much of the full disc scale (30$'$) emission. For a dynamical structure like the Sun, temporal resolution is just as vital, with chromospheric evolution sometimes happening on timescales shorter than a second \citep{Kontar2018}.

\subsection{Transients and Variability}
\label{SciTheme_Transients}
\textit{See \citet{Orlowski-Scherer2024} for further details.}

As can already be seen in some of the science cases presented above, another dimension has emerged as important in understanding the universe as it evolves around us: time. The study of transient and variable phenomena is still in its early stages in the sub-mm regime, with the first papers on the subject published less than 10 yrs ago, but already we are seeing evidence for variability in objects from AGN and Gamma-Ray Bursts (GRBs), through to supernovae and forming stars and the atmospheres of the Sun and planets. Transient and variability studies can help shape theoretical predictions for a wide range of astrophysical phenomena, but care must be taken to ensure the proper types of observations (e.g montoring, triggered ToOs and serendipitous discoveries), and on the right cadence, for time domain astronomy to come to prominence. 

Some transient detections have come from the sub-mm \citep{Guns2021, Lee+21, Li2023}, however most are followups of detections at other wavelengths.  With the right facility, software and operations model, AtLAST could instead be the source of transient detections which will require followup and monitoring by both AtLAST and other facilities. Take, for example, the JCMT transient study of nearby star forming regions \citep{Herzceg+2017,Mairs+17,Mairs+2024}. These studies have shown the variability of local protostars, and the observatory is continuing to monitor the same regions to detect longer periods of variability.  However, current facilities are hampered by small fields of view and sensitivity. The former limiting how many targets can be monitored, the latter limiting how precise the measurements can be as well as their overall cadence.  

Variability can happen on timescales of minutes to decades. Observing strategies need to be in place to ensure that proper monitoring cadences are maintained. In addition to planned observations of variable sources, the high sensitivity provided by a large aperture telescope greatly increases the chances of serendipitous detections found in datasets intended for other purposes. To ensure that these are picked up on short timescales to enable the triggering of follow-up observations, real-time data reduction pipelines for transient detection are also required.  
These types of commensal observations, allow for additional science goals to be met using observations designed often for a very different purpose. Many of the upcoming CMB experiments are expected to have such data processing software in place, which will already start taking stock of the transient sky. With 100x the resolution, and significantly better sensitivity, AtLAST will be able to detect more transients that would get lost in the large beams of the CMB surveys.

Following up on transient and variable phenomena will necessitate observing modes like target of opportunity (ToO) overrides, coordinated observations with other facilities for multi-wavelength observations, monitoring campaigns and the commensal observing described above.

\subsection{Black Hole Event Horizons}
\label{SciTheme_VLBI}
\textit{See \citet{Akiyama2023} for further details.}


The global VLBI network has played a significant role in our understanding of black hole physics, showing the emission from lensed material very close to the black hole itself for the first time \citep{EHTM87PaperI}. The existence, nature and shape of this emission in M87 and SgrA$^*$ definitively proved the existence of black holes at the centre of galaxies and as the power sources for relativistic jets; provided unique probes of the astrophysics of accretion and relativistic jet formation; and enabled the exploration of properties of general relativity.

By increasing the network, the EHT will be able to image ever fainter sources, particularly with the addition of a telescope as sensitive as AtLAST. Adding a sensitive anchor not only improves the baselines that AtLAST is a part of, but also those between other stations due to improvements in calibration. As a single dish, AtLAST has major advantages over ALMA in this regard as ALMA has to act as a phased array, which limits the sensitivity of sources it can detect. AtLAST will also be able to host an instrument with multi-frequency capabilities, enabling both simultaneous multi-frequency observations \citep[e.g.][]{Issaoun2023} and order of magnitude improvements in sensitivity due to the frequency phase transfer technique \citep{Akiyama2023}.

The images of M87 and SgrA$^*$ are right at the resolution limits of a 230 GHz global VLBI network, yet, within those rings are an infinitely nested set of separate rings \citep{Johnson2020}, each of which reveals new insights into the nature of the black holes powering them. Shifting to higher frequencies (690~GHz) will allow future EHT observations to resolve the n=0 and n=1 rings of these black holes (at $\sim$ 6 microarcsec) and disentangle the astrophysical and relativistic effects in that emission \citep[e.g.][]{Tiede2022}.

\clearpage

\section{Telescope Requirements and Generalised Instrumentation}
\color{black}
\label{sec:requirements}

\begin{figure}[ht]
    \centering
    \includegraphics[width=0.9\textwidth]{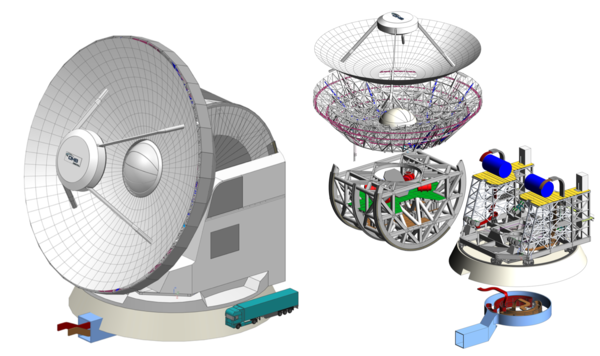}
    \caption[] 
    {AtLAST computer aided design model. \citep[Figure reproduced from][]{Mroczkowski2024}}
    \label{fig:cad}
\end{figure}

The science cases presented in the previous section place various requirements on the telescope, instrumentation and operations that we now discuss here. A summary of the key requirements taken from each science theme is shown in Table \ref{tab:requirements_overview}, with additional telescope requirements listed here which are implicit in the science cases. Many of the telescope requirements described below are already incorporated into the first designs of the telescope, as described in \citep{Mroczkowski2024,Kiselev2024,Gallardo2024,Puddu2024}, and shown in Figure \ref{fig:cad}.

\subsection{Telescope requirements}
\label{sec:tel_requirements}

\subsubsection{Single dish}

The large scale mapping described in many of the science cases requires the use of single dish facilities because interferometers suffer from a lack of sensitivity to large-scale emission; the size scales on which they start to become insensitive typically being described as the maximum recoverable scale (MRS). For ALMA in its compact (low resolution) configurations at sub-mm wavelengths, this scale is of order a few arcseconds, much smaller than the largest angular scales of many of the targets considered here, which range from arcminutes (e.g. comet halos and debris discs) to degrees (e.g. GMCs within our own galaxy, nearby galaxies, the CGM, galaxy clusters).  This problem can even impact scales smaller than the MRS \citep{Plunkett+23}, in both spectroscopy and imaging \citep{Hacar+18}. For extended targets, a single-dish telescope is necessary to ensure an accurate measure of the flux and avoid any biases introduced by interferometric artifacts.

The stability in resolution offered by a single dish telescope is ideal for consistent observations of time critical objects and those that require cadenced monitoring. Interferometer arrays are often built to allow changes in configuration, which  is beneficial for enabling a range of angular resolutions. The changes in configuration are typically time consuming, and result in the arrays staying in a single configuration for a pre-determined amount of time (e.g. ALMA spends about a month in each configuration). This is problematic for time critical observations where the array may not be able to provide the appropriate resolution at the requested time of the observations. Seasonal monitoring of planetary atmospheres and the Sun are examples of such cases that need a consistent resolution for periodic observations throughout the year.

\subsubsection{Dish diameter}
\label{require_50m}

The dish diameter impacts both the angular resolution ($\approx1.2\lambda/D$) and sensitivity ($\propto1/D^2$) of observations. There is a fundamental limit to how sensitive an observation an observatory can make, and that is related to the so-called `confusion noise' limit of the observatory, which is directly related to its angular resolution.  As described more in Appendix \ref{app:confusion}, as observations become more and more sensitive, fainter or more distant objects begin to show up within an observed field of view. There comes a point where there are so many objects detected that they can no longer be distinguished from each other; they become confused within a single pixel.  For higher resolution observations, the confusion limit is lower, which means we can push further.  Current and planned observatories capable of observing at sub-mm wavelengths reach confusion limits outside the galactic plane that mean they cannot observe the `normal' population of galaxies, only resolving the brightest ones.   In order to detect galaxies of luminosity L$^*$ out to $z=5$ and resolve 97\% of the CIB, a 50~m dish is necessary. With a dish of only 40~m, the confusion limit would be twice as high, 15\% less of the CIB would be resolved and our ability to reach L$^*$ would only reach as far as $z=4$. This confusion limit due to high redshift galaxies is important not only for studying the galaxies themselves, but also for faint foreground targets, such as nearby galaxies (both in terms of their ISM and CGM) and debris discs as faint as the Kuiper belt around the closest ($<$10~pc) stars, where sensitivities on the order of 10s of $\mu$Jy/beam at sub-mm wavelengths are necessary, which is well below the confusion limit of current single-dish sub-mm facilities.

A small primary beam ($\lesssim8''$ at 1.1~mm, corresponding to $\gtrsim$34m dish diameter) is strictly required for observations of Saturn's moon Enceladus, to avoid contamination and saturation from its close proximity ($<25''$) with the bright emission sources of Saturn and its rings. The planets themselves can all be spatially resolved with a primary beam $\lesssim2''$ (possible with a 50m dish at 690 GHz), which enables detailed atmospheric studies.

To disentangle star forming regions from their environments, and study their interactions, requires the resolution to identify star forming clumps (1-0.2~pc) and cores (0.1 pc structures) within their natal clouds across the plane of our Galaxy. With a 50m diameter, those resolutions can be achieved at the wavelengths best suited to observing those structures out to 10 kpc (i.e. beyond the central molecular zone): in low frequency N$_2$H$^+$ and HCO$^+$  transitions at $<$ 100 GHz we will achieve a resolution of $\sim15''$ (0.7 pc at 10 kpc) towards cold star forming clumps, and in higher frequency CO, CS and other species at $>$ 600 GHz we will achieve a resolution of $\sim2''$ (0.1~pc at 10~kpc) to probe the hotter and more chemically rich star forming cores.  Further afield, a 50~m primary allows us to resolve GMCs ($\sim$300 pc in scale) out to 20 Mpc at 460 GHz (where we achieve $\sim3''$ resolution).

In addition to these science drivers is the size required for combination with ALMA data. In order to ensure overlapping coverage in the Fourier plane, a single-dish telescope is required to have a diameter of at least      $D_{\rm{SD}}\geq1.18L_{\rm{min}}/\kappa$ \citep{Frayer2017} where $L_{\rm{min}}$ is the minimum baseline \citep[14.6~m for the compact ALMA main array configurations,][]{Cortes2023} and $\kappa$ is a coefficient between 0.4 and 1, depending on the source distribution and signal-to-noise of the data. If we conservatively take this to be 0.4 in order to cover edge cases, then we find that a dish size of at least 43~m is necessary to facilitate full coverage of the short baselines when combining with ALMA's main array.

\subsubsection{Field of view}
\label{require_fov}
Many of the key science cases described in this document require surveys with a large spatial coverage either because they target large numbers of sources or sources with a large angular scale. Surveys of high redshift galaxies fall into the former category, with an expectation of finding around 50,000 galaxies per $1~\mathrm{deg^{2}}$ field in continuum observations and a desire to cover $\geq1000~\mathrm{deg^{2}}$. Whereas studies of the plane of our own galaxy are expected to cover up to 540\,deg$^2$ (covering the full 270$^{\circ}$ plane visible from the southern hemisphere, with a degree above and below the midplane), the CGM around nearby galaxies extends to a few degrees, SZ effects have scales on the order of a degree, and the Sun is half a degree across. Observing these large scale structures efficiently requires a large field of view filled with detectors. In general, a  larger \textit{filled} FoV reduces the time needed for surveys of such targets.  It also necessitates calibration strategies that do not lose that emission due to atmospheric fluctuations (see next section); in the SZ case an instantaneous field of view covering $>1~\mathrm{deg^{2}}$ is explicitly needed as a smaller FoV will not have sensitivity to the degree scale structure essential for the project. Transient and variable targets can appear anywhere on the sky and so a larger field of view increases the chance of detecting such phenomena, especially in cases where they are poorly localised triggered events, or are commensal (i.e. transients to be detected by automated routines in observations derived from other science goals).

\subsubsection{Observing at Sub-mm wavelengths}
\label{require_submm}
\begin{figure*}
\centering
   \includegraphics[clip=true,trim=0.2cm 1.6cm 0.2cm 3.4cm,width=.95\textwidth]{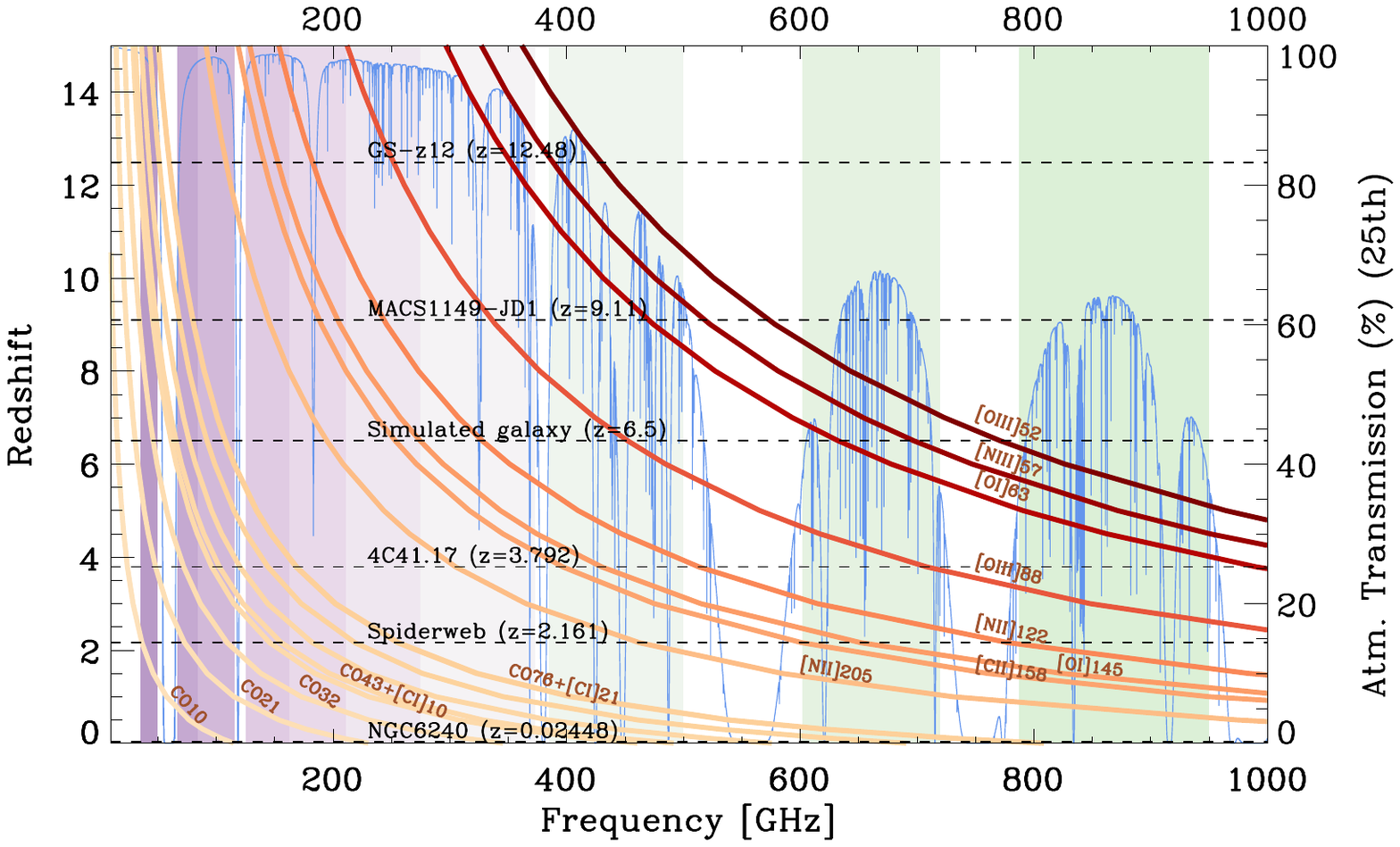}
      \caption{Evolution of the observed frequency of the brightest sub-mm and far-infrared emission lines from CGM (orange/brown lines) as they get redshifted into the sub-mm bands at different redshift values. The blue spectrum in the background shows the top quartile of the atmospheric transmission at the Chajnantor Plateau (about 5100 metres above sea level, derived using the atmospheric \texttt{am} code, \citealt{Paine2019}), where the corresponding transmission is reported on the y-axis on the right. The coloured vertical bands correspond to the ALMA bands (see Appendix \ref{app:freq_bands}). The horizontal dashed lines indicate the redshifts of a few sources discussed in the paper, and of two additional high-z galaxies from the literature \citep{D'Eugenio2023, Hashimoto2018}. This plot demonstrates the importance of a broad wavelength coverage for the CGM science case, and it highlights how crucial the high-frequency bands are to observe the brightest tracers of the cold CGM in galaxies at $z>1$. \citep[Figure reproduced from][]{Lee2024}}
   \label{fig:plot_bands_lines}
\end{figure*}

Cold dust in the Universe has a peak rest-frame wavelength around 100~$\mu$m, which is shifted to longer wavelengths at higher redshifts. The sub-mm atmospheric windows therefore provide the only way to probe this emission at or close to the peak of the distribution at high redshift from a ground based observatory.   Figure \ref{fig:plot_bands_lines} also shows key atomic and molecular line frequencies as they shift into the sub-mm bands as a function of redshift. To observe the CGM of galaxies at z$<1$, emission too large scale and too diffuse to detect with current generation facilities, requires the highest sub-mm wavelengths.


The sub-mm atmospheric windows open up the possibility of observations of redshifted spectral lines. For instance, being able to detect multiple tracers of the CGM in the same source enables the study of the physical properties of the gas, its density, temperature, metal content, mass, and so its origin. Given the wide range of redshifts (from $z=0$ to $z=10$) that are of interest for the CGM science case and studies of distant galaxies in general, a telescope with sub-mm (e.g. $\lambda < 600~\mu$m) capabilities would have access to the brightest gas tracers of material in a larger volume of the Universe (see Figure \ref{fig:plot_bands_lines}). For observations within our own galaxy, many rotational transitions of important molecules emit at sub-mm wavelengths; some tracing the thermal balance of the gas, others tracing its ionisation state or the dynamics of the warmer gas traced by the higher energy rotational transitions of molecules.  For comets in the Solar System, access to the HDO $1_{1,1}-0_{0,0}$ line at 894~GHz is useful to the study of D/H ratios that can inform us of the origins of Earth's water.

For these reasons, AtLAST needs to be located in a high dry site and needs to have a high surface accuracy to access the sub-mm window.

\subsubsection{Scanning speed}

Follow-up of some transient sources that decay rapidly such as GRBs or gravitational wave events requires rapid slewing to quickly get on target and begin observing. A fast scanning speed is also necessary for calibrating out atmospheric fluctuations and mapping large fields.  These needs push for a scanning speed of a few degrees per second, with one of the most stringent requirements set by the need to detect faint emission on large scales that would be drowned out by short term atmospheric fluctuations if not accounted for properly; the ability to modulate the astrophysical signal faster than the signal from Earth's atmosphere \citep{Morris2022, vanMarrewijk2024b}. For AtLAST, an average scan speed of 1.5~deg/s is needed to keep the atmospheric signal stable in the time domain, with the ability to reach 3~deg/s needed for faster than average wind conditions \citep{Mroczkowski2024}.

\subsubsection{Pointing and Tracking}

With many of the science cases described above, large scale mapping is key. To ensure these maps are of high quality requires high relative and absolute pointing calibrations (greater than the resolution of the telescope) and tracking ability.  

Ensuring proper pointing is also important for instruments with greater than Nyquist spacings between their individual pixels which means that small maps (i.e. jiggle, dither, etc) are required to fully sample the sky.

\subsubsection{Beam accuracy}

Accurate calibration is needed to properly account for the potential systematics in the small-amplitude fluctuations of the SZ signal that are associated with local pressure and velocity perturbations, or to relativistic distortions. This results in a requirement of a sub-percent level control of the beam stability. This is also important for cases that need a high dynamic range such as observing moons near bright planets, planetary atmosphere observations that need to distinguish the weak atmospheric lines from the bright continuum or observations of the CGM, which is much fainter than the ISM.

\subsection{Instrumentation requirements}

The science cases in this report require a mix of continuum and spectral line observations including polarisation, solar observing and VLBI capabilities. 

\subsubsection{Continuum Observations}

Cold dust throughout the Universe (from galaxies to circumstellar discs to asteroids within our own Solar System), photons scattered by hot electrons due to the SZ effect and emission from the atmosphere of our own Sun can all be observed in their broadband continuum emission or as deviations from that expected continuum. A wide wavelength coverage from 350~$\mu$m to 3~mm is desirable to get a broad sampling of the spectral energy distribution (SED) of the targets. Sub-mm observations, in particular, are crucial as these wavelengths are closest to the peaks in the SEDs for many targets and due to the improvements in resolution that they provide (see Section \ref{require_submm} for more details). These continuum emission measurements  often need to be combined with observations at other wavelengths for the greatest scientific benefits. For instance, when considering thermal emission from dust, multi-wavelength observations enable information on the dust composition and size distribution to be derived. At around 1~mm and longer, other contributions to the SED, such as free-free emission, can dominate and so multi-wavelength observations allow these distinct contributors to the overall continuum emission to be disentangled. Multi-chroic observations are also important for distinguishing between the thermal and kinetic SZ effects. 

A large number of the science cases described above would benefit from simultaneous multi-chroic (or multi-band) observations (see, for instance figure \ref{fig:survey1000}).  For transient and variable sources, simultaneity reduces the need to distinguish between calibration and temporal effects, and in the case of short timescale phenomena such as solar flares, it is necessary for multiple wavelengths to be observed simultaneously to capture the evolution of the flare in real-time.   However, for deep observations (such as high redshift galaxies or extrasolar Kuiper belts), multi-chroic observations may be less efficient as the longer wavelength detectors will become confusion limited much quicker than the shorter wavelengths (see Appendix \ref{app:confusion}).

The bandwidth of the continuum camera is a key parameter for determining the sensitivity of an observation, with wider bandwidths allowing for greater sensitivity in a given integration time. To maximise observing efficiency, getting bandwidths as close to the (sub-)mm atmospheric transmission windows as possible would minimise the on source integration times (see Appendix \ref{app:freq_bands} for the bounds of those atmospheric windows).

\begin{figure}[hbt]
    \centering
    \includegraphics[width=1\textwidth]{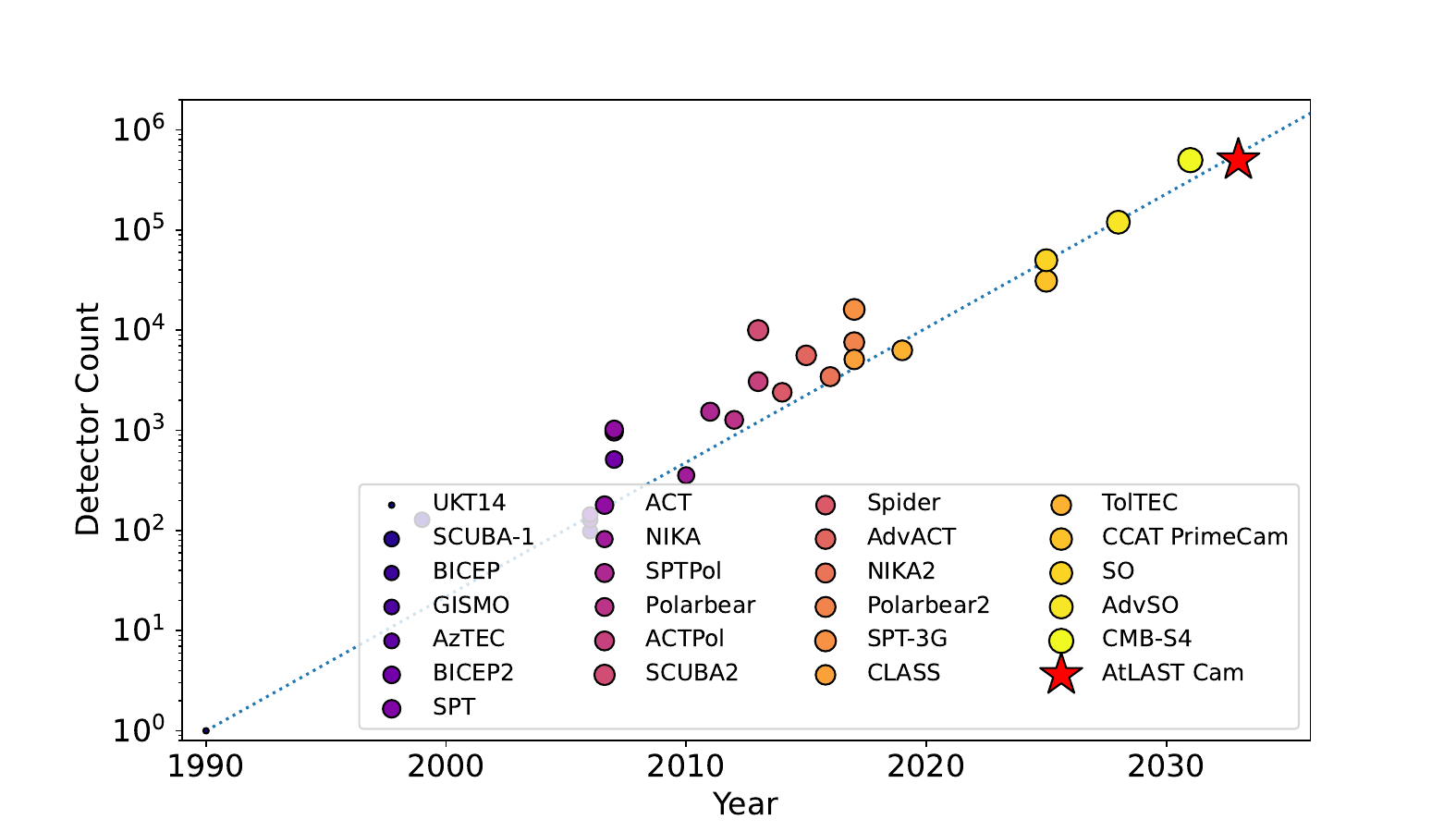}
    \caption{Number of pixels in existing and expected continuum cameras on current and planned facilities as a function of time.  Extrapolating from the trend seen in this plot, is is expected that a first light continuum camera for AtLAST could have upwards of 10$^6$ pixels (as shown with red star in the top right corner of the plot).}
    \label{fig:nbolo}
\end{figure}

As described in Section \ref{require_fov}, a FoV of $>1~\mathrm{deg^{2}}$ is required for many of the science cases due to either the large scale emission from the targets (e.g. CGM and the Galactic plane) or the large number of targets (e.g. surveys of distant galaxies) available to be observed. To properly take advantage of that large field of view requires highly multiplexed instrumentation.  
This opens up significant possibilities for synergy between various science cases. 
For example, large area (1000+~deg$^{2}$) continuum surveys can be used to study distant galaxies and the clusters and protocluseters they inhabit, whilst also picking up any nearby circumstellar discs within the field as well as serendipitously detecting any transients or variables. A 300\,000 pixel camera would be capable of a 2.2 degree FoV at 3~mm or provide a 0.26 degree FoV at 350~$\mu$m (if the pixels are well spaced). However, this assumes one camera per wavelength. As some of the cases push for a multi-chroic camera, this could result in trade-offs between FoV and number of wavebands if the total number of pixels that can be placed in an instrument is fixed (see, for instance, Figure \ref{fig:nbolo}). 

\paragraph{Solar Observing} 

Observations of the Sun also place requirements on the detectors as it is the brightest mm source in the sky and can vary in brightness significantly, in particular during flares. A solar filter with well-defined characteristics and large dynamic range for the detector are needed and may be best served by a specialised solar observing instrument. 
While integration times are very short for a bright object like the Sun, spatial (or spectral) scanning in order to cover an adequate spatial (or spectral) region is constrained by the short dynamic timescales. The temporal cadence should be at most 1\,min and ideally less than 1\,s as flares develop quickly. High detector readout rates are required, especially for fast scanning. Given the particular requirements of solar observing, a bespoke instrument may be better suited to fulfilling this requirement.

\subsubsection{Spectral Line Observations}

Atomic, ionised and molecular gas fills the ISM of galaxies, extends beyond them as part of their CGM and even populates the ICM of clusters. Studying the spectral signatures of this gas tells us about the chemistry, temperature and ionisation state within the gas and its dynamics both relative to other parts of the galaxy and relative to us -- allowing us to determine redshifts. Many Solar System bodies are surrounded by atmospheres of gas, which is rich in information about their origins and ongoing chemistry.

\begin{figure*}[t]
    \centering
    \includegraphics[width=0.98\textwidth]{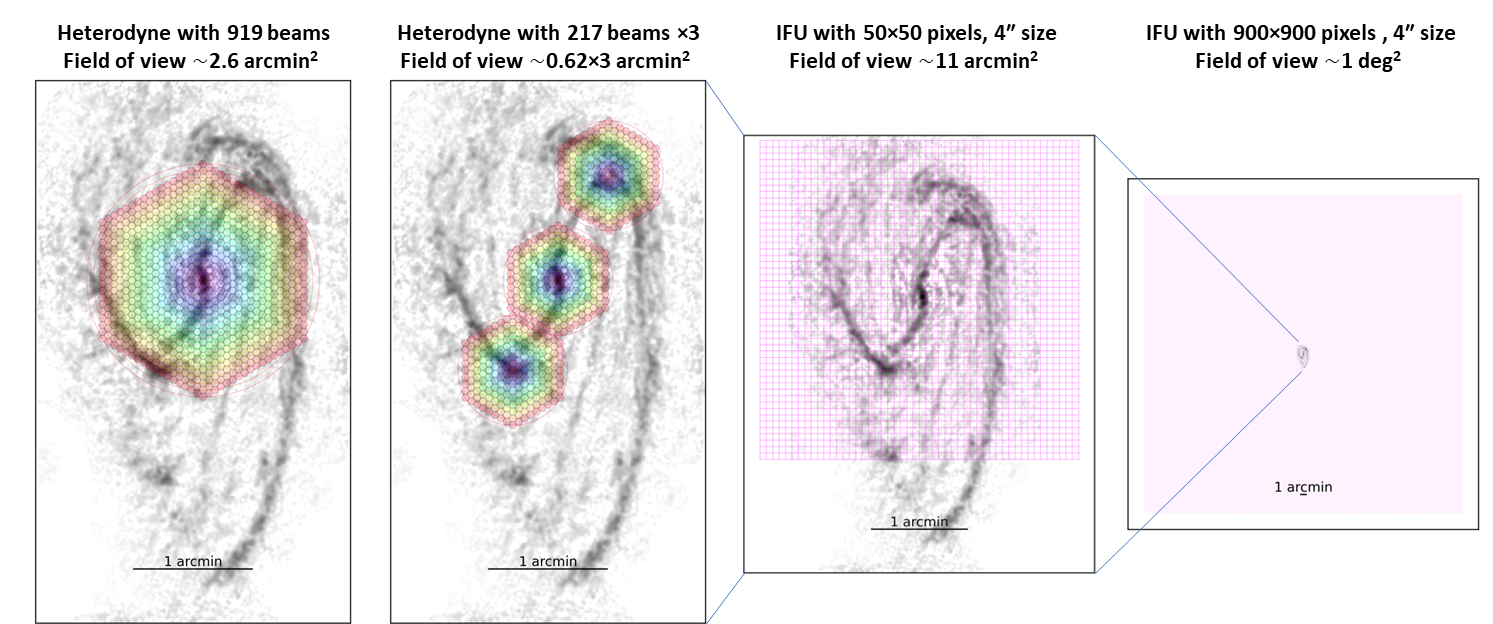}
    \caption{%
    Footprints of various assumed instruments (from left to right): a 919-beam (18-layer) heterodyne receiver, a 217-beam (9-layer) heterodyne receiver, an IFU with $50 \times 50$ pixels, and an IFU with $900 \times 900$ pixels, on top of the nearby galaxy NGC~3627's CO(2--1) emission \citep{Leroy2021b}. 
    The heterodyne receiver's hexagon units and the IFU pixels are assumed to have a spacing/size of $4''$, which is about the AtLAST 50-m angular resolution at 400~GHz. 
    A large-format IFU as shown in the last panel can most efficiently observe multiple galaxies, with each galaxy fully sampled at resolutions as shown in the first three panels. \citep[Figure reproduced from][]{Liu2024} 
    \label{fig:heterodyne}
    }
\end{figure*}

Large format spectrometers are necessary to observe the gas populations described above and quantify ionisation states, dynamics, chemistry, evolution and redshift (see, for instance figure \ref{fig:heterodyne}). Highly multiplexed receivers are key to any spectral science cases that aim to study wide areas, but different types of spectroscopic instruments are better suited to the broad range of these science cases. For many Galactic and nearby galaxy science cases, high spectral resolution (of order $<0.1$ km s$^{-1}$, or R $>10^8$) is required to disentangle kinematics and hyperfine transitions of molecules that have linewidths below 1 km s$^{-1}$. Instrument design, processing power and data transfer rates then limit the bandwidth that can be simultaneously observed with such an instrument. For nearby galaxies, with much broader spectral lines (of order a few 100 km s$^{-1}$), the same instrument could be used, likely in a lower spectral resolution mode, but with a greater need for stable spectroscopic baselines to ensure detection of faint, non-Gaussian, emission in the line wings. To ensure these baselines would require position switched or other atmospheric corrective observing techniques. This is contrasted against science cases dedicated to finding the redshifts of galaxies within the FoV of the instrument where much lower spectroscopic resolution is required (R$\sim$ few hundred). Because these science cases can get by with lower spectral resolution integral field units (IFUs) they can simultaneously observe much larger spectral bandwidths which enables covering larger redshift ranges, which increases the efficiency of such `redshift machines'.

Based on current technologies, and extrapolating to future spectroscopic instruments \citep[see, for instance][]{2019Groppi}, 1000 pixel spectroscopic instruments will become feasible by the time AtLAST has first light.

\subsubsection{Polarimetry}

Polarimetric observations are important for understanding the interplay between the various forces (i.e. gravity, turbulence, magnetic fields) at work in the clouds and galaxies and clusters under study.   Such capabilities for the continuum and spectral cameras are a key requirement unlocking a hidden dimension of understanding, particularly for studies of magnetic fields in galaxies (including the Milky Way) and our Sun.  Continuum polarimetry (as shown in the left panel of Figure \ref{fig:cra}) gives an understanding of the directionality of the dust grains in the object(s) being observed.  Spectropolarimetry preserves the polarisation information present in molecular lines, and thus can differ from that in the dust continuum.

To achieve polarimetric observations not only requires polarisation preserving instruments, but optical paths which do not confuse the polarisation signals.  The former can often be integrated into MKID and other detector-on-chip type instrument technologies, while the latter needs to be incorporated into the observatory at conceptual design time \cite[see, for instance,][]{Puddu2024}.

\subsubsection{Very Long Baseline Interferometry}
The incorporation of a large, single-dish telescope capable of sub-mm observations into EHT will significantly increase the number of viable targets and the science return of the Event Horizon Telescope. For AtLAST to become a VLBI station, a specialised heterodyne receiver is necessary that is capable of high sensitivity observations at multiple frequencies (at least 90, 230, 345 and 690~GHz) with a wide bandwidth.

\subsection{Telescope Operations}
To address the ambitious and diverse scientific programme outlined in this report, and to accommodate future user-driven research initiatives, a versatile and sustainable operational scheme must be developed. This range of programmes requires a flexible scheduling system with suitable observing modes. Additionally, the powerful capabilities of AtLAST, such as its large field of view, high sensitivity and multi-beam and multi-pixel instrumentation, necessitate efficient methods for processing, storing, and providing the community with high-quality data for optimal exploitation.

\subsubsection{Observing modes}
The variety of science cases require different types of observing strategies. Telescope scheduling must accommodate requirements from large, automated surveys of the sky to dedicated principal investigator (PI) projects requiring only a few hours on sky, and everything in between. This diversity in observing modes calls for fully dynamic scheduling to quickly adapt to the weather conditions and to make the best use of the available time, allowing for flexibility and on-the-fly changes of the schedule.

PI-driven programmes can range from a few to a few hundreds of hours (in the form of Large Programmes), depending on the nature of the sources to be observed and scientific goals being pursued. Depending on the case, granting the PI the possibility to interact with the observatory staff, to check the quality of the data while it is being taken and to refine the observing strategy, when necessary, will help maximise the scientific output of the facility.

To better meet the requirements for Solar studies, a combination of long-term synoptic observations with a very low daily load on observing time with campaign-based observations is needed. The observations of active regions should be schedulable with just a few days warning, as the coordination with other space-borne and ground-based observatories will optimise the scientific result.

Transient and variable phenomena from the Sun and other Solar System objects to GRBs and AGN are time critical and require special observing modes to be considered. Target-of-opportunity observations, coordination with other telescopes/space missions, and accommodation for regular monitoring of science targets with a predefined cadence (over a timescale of hours, days, weeks, months or years), will all be crucial for maximum science return on time-variable sources. Scheduling the most critical observations to within an accuracy of a few minutes will therefore be desirable. Responding to unexpected transient phenomena such as cometary outbursts or gravitational wave events will necessitate flexible telescope scheduling, for example, with turnaround times from a few minutes to a few days.

If AtLAST is to not only follow-up transient triggers from other telescopes but also to send its own triggers when transients are serendipitously detected, then a transient search pipeline with an automated alerting system should be set up. This can be implemented with an ``opt-out'' option for PIs in cases where their science goals include searches for transients.

Coordination with other telescopes will also be necessary for VLBI observations. This places extra constraints on the operations with specialised equipment in addition to the heterodyne instrumentation including a highly phase-stable frequency standard, backend digital electronics for recording data at rates up to 256 Gbps and specialised control software.

\subsubsection{Data flow and legacy value}
AtLAST is expected to have a tremendous legacy value given the variety of different imaging and spectroscopic surveys the facility will address. The commitment is to enhance this by providing users with fast, reliable and useful access to AtLAST data, as well as maintaining high data quality standards. This requires supplying an adequate infrastructure to store the data and give easy access to the community. The final goal is to make data accessible, properly described and compatible with other analysis tools and services that facilitate comprehensive research across wavelengths, in line with the FAIR principles \citep{Wilkinson2016}.

The large format continuum and spectral line cameras onboard AtLAST will generate large data volumes, especially if the spectral line cameras can achieve a 0.01 km/s spectral resolution across a 32 GHz bandwidth. This will require re-thinking calibration and data reduction techniques and implementing more efficient and sustainable data processing and storage schemes.

AtLAST will not be alone with this problem. With ALMA’s Wideband Sensitivity Upgrade \citep{Carpenter+2023}, and the even larger fields of view possible with the SKAO, data storage, reduction and analysis will become significantly more challenging. This will open new opportunities, on the one hand to explore and incorporate new technological developments, like novel distributed computing systems or federated cloud services, and on the other hand establishing synergies among different research infrastructures facing these common challenges.

\clearpage
\section{Summary}
\color{black}

In this document, we have summarised the key science cases developed for AtLAST by the community through the consultation process held during the AtLAST design study.  The community consolidated around 28 use case submissions in the first part of the consultation \citep[as summarised in][]{Ramasawmy2022} and further delved into the details of these science cases in five working groups which produced the 7 science white papers summarised in Section \ref{sec:themes}. The technical requirements these science cases place on the observatory were then discussed in Section \ref{sec:requirements}, with the key science drivers for AtLAST, derived from these consultations presented in Table \ref{tab:key_questions}. An overview of the technical constraints from the overarching science themes are consolidated into a matrix of requirements in Table \ref{tab:requirements_overview}. 

The strong support of the community has been vital to this work, with experts in various (sub-)mm domains volunteering their effort towards understanding the technical needs of future facilities in order to advance our understanding of the Universe around us.  Similarly, the thoughts and views and priorities of this community will evolve along with our understanding. As such, this document should be viewed as a snapshot of the anticipated science return that an observatory like AtLAST could provide. As new and better observatories and instruments come online, models improve and science evolves, so to will the telescope requirements for enabling great science.  The future evolution of the telescope design must take this into account and continue refining its science goals in the light of new discoveries.

\NewDocumentCommand{\rot}{O{60} O{1em} m}{\makebox[#2][l]{\rotatebox{#1}{#3}}}%

\begin{table}[hbt!]
    \centering
    \addtolength{\tabcolsep}{-0.1em}
    \begin{tabular}{p{0.05cm}l|ccccccccc}

     &    & \rot{The Sun} & \rot{Solar System} & \rot{Milky Way} & \rot{Nearby Galaxies} & \rot{CGM} & \rot{Distant Universe} & \rot{The SZ effect} & \rot{VLBI} & \rot{Transients} \\
    \hline
   \multicolumn{2}{l|}{\textit{Telescope}}&\\
    &Diameter (m) & $>$50 & $>$34 & $>$50 & $>$53 & $>$50 & $>$50 & $>$43 & $>$50 & $>$50\\
    &Field of View (deg) & 0.5 &0.2 & $>$1 & 1 & $>$1 & $>$1 & $>$1 &- & $>$1\\
   & LAS$^{a}$ (deg) & 0.5 & 0.2 & 10 & 70 & 1 & - & 1 & - & -\\
    & Angular resolution$^{b}$ ($''$)&$<$5&$<$7&$<$5 &$<4.5$ &$<$5 & $<$5&$<6$ & (6e-6)$^{d}$ &- \\
    &Spatial Coverage (deg$^2$) & 0.25 & 70.5 &  540 & 1000+ & 1000+&1000+&1000+&- & 540 \\
   \multicolumn{2}{l|}{\textit{Wavelengths}}&\\
    &Atmospheric Windows$^{c}$ & 1-10 & 3-10 & 3-10 & 3-10 & 1-10 & 1-10 & 1-10 & 3-9 & 1-10 \\
   \multicolumn{2}{l|}{\textit{Spectral Requirements}}&\\
    &Spectral observations  && \checkmark{}& \checkmark{} & \checkmark{} &\checkmark{}& \checkmark{}&&\checkmark{}&\checkmark{} \\
    &Polarisation  && & \checkmark{} &  && \checkmark{} && \checkmark{}& \checkmark{}\\
    &Spec. Bandwidth (GHz) & - & 64 & 16-32 & 32 & 4-8 &400 & -& $>$8& -$^e$\\
    &Spec. Resolution (km/s) & - & 0.1 & 0.01 & 0.5 & 5 & 400 & -&-&-$^e$\\
    &Spatial Sampling &Full &Full & Full/MOS & Full/MOS & Full&Full/MOS&-&-&-$^e$ \\
   \multicolumn{2}{l|}{\textit{Continuum Requirements}}&\\
     &Bandwidth  & \multicolumn{9}{c}{Full atmospheric window}\\
     &Resolution   & \multicolumn{9}{c}{1+ channel per band} \\
     &Spatial Sampling &Full &Full & Full & Full & Full&Full&Full&-&Full \\
   &Polarisation  &\checkmark{}& & \checkmark{} & \checkmark{} && &&&\checkmark{} \\
    \multicolumn{2}{l|}{\textit{Timing}}&\\
    &Repeats &\checkmark{} &\checkmark{} & \checkmark{} & \checkmark{} &&  &&&\checkmark{}\\
    &Time Variability  & \checkmark{}& \checkmark{}&\checkmark{}  &\checkmark{} & &&&&\checkmark{}\\
    &Targets of Opportunity & \checkmark{}&\checkmark{}  &\checkmark{}& \checkmark{} &&&&&\checkmark{}\\
    &Cadenced observations  &\checkmark{}& \checkmark{} &\checkmark{}  &\checkmark{} & & &&&\checkmark{}\\
    &Time Critical &\checkmark{}& \checkmark{} &\checkmark{} &\checkmark{}&&&&&\checkmark{}\\
    
    \hline\hline
    \end{tabular}
    \caption{Generalised telescope and instrument requirements defined by the science working groups. $^{a}$Largest angular scale. $^{b}$Angular resolution at 1~mm. $^{c}$The Atmospheric windows are listed with respect to the ALMA observing bands, for which a translation to wavelength/frequency is found in Appendix \ref{app:freq_bands}. $^{d}$ Resolution of VLBI observations not set by AtLAST alone. $^e$ Spectral and spatial sampling requirements vary according to the type of variable target being observed within the other science categories presented here.}
    \label{tab:requirements_overview}
\end{table}

\appendix
\addtocontents{toc}{\protect\setcounter{tocdepth}{1}}

\section{Frequency Ranges of `ALMA Bands'}
\label{app:freq_bands}

The current and anticipated receivers on the ALMA antennas are well aligned to the (sub-)mm atmospheric windows.  The values in Table \ref{tab:ALMABands} describe the rough edges of the frequency bands defined within the ALMA community, and correspond to the vertical bands presented in Figure \ref{fig:plot_bands_lines}.

\begin{table}[hbt]
    \centering
    \begin{tabular}{c|cc||c|cc}
    \hline \hline
         ALMA Band  & Wavelength & Frequency & ALMA Band  & Wavelength & Frequency \\
                    & (mm)       & (GHz)   & & (mm)       & (GHz)\\
        \hline
         1          & 6-8.5     & 35-50 &        6          & 1.1-1.4   & 211-275 \\
          2          & 3.3-4.5   & 65-90 &  7          & 0.8-1.1   & 275-373\\
         3          & 2.6-3.6   & 84-116 & 8          & 0.6-0.8   & 385-500\\
         4          & 1.8-2.4   & 125-163 & 9          & 0.4-0.5   & 602-720 \\
         5          & 1.4-1.8   & 163-211  &          10         & 0.3-0.4   & 787-950 \\
 
         \hline \hline
    \end{tabular}
    \caption{Mapping of ALMA Observing bands to their Atmospheric windows}
    \label{tab:ALMABands}
\end{table}

\section{Confusion Limit}
\label{app:confusion}

`Confusion' refers to the issue of lacking the angular resolution necessary to spatially distinguish between sources on the sky that are physically separate. 
The deeper an observation, the more sources are detected and the worse this problem becomes leading to a limit on the sensitivity which is referred to as the confusion limit. 
Confusion can come from both Galactic and extragalactic sources but in the \submm{}, the main source of confusion is distant galaxies \citep[e.g.][]{Blain2002}. 
There are then multiple criteria that can be used to define the confusion limit \citep[e.g.][]{Dole2003,Ermash2020}.
Here we consider both the photometric criterion, which is most relevant when considering the confusion as a source of background noise, and the source density criterion, which is most relevant when studying the galaxies themselves. 

The confusion defined by the so-called photometric criterion corresponds to signal fluctuations in the beam of the observation due to extragalactic sources below the detection threshold $S_{\rm lim}$.
Following \citet[][]{Dole2003}, it is given by
\begin{equation}
\label{eq:conf photo 1}
\sigma_{\rm c} = \int_\Omega f^2(\theta,\phi)\,d\theta d\phi \int_0^{S_{\rm lim}^{\rm c}} S^2 \frac{dN}{dS} dS,
\end{equation}
where $f(\theta,\phi)$ is the two-dimensional beam profile, $S$ is the source flux density, $dN/dS$ is the differential number counts, $\sigma_{\rm c}$ is the photometric confusion noise and $S_{\rm lim}^{\rm c}$ the photometric confusion limit.
These last two variables are then implicitly linked by
\begin{equation}
\label{eq:conf photo 2}
S_{\rm lim}^{\rm c} = q_{\rm phot} \times \sigma_{\rm c},
\end{equation}
where $q_{\rm phot}$ measures the desired photometric accuracy and is usually taken equal to 5.
As the calculation of this photometric confusion limit relies on the knowledge of differential number counts down to infinitely low flux densities, it requires the use of a galaxy population model.
Here we used the empirically-motivated galaxy population model of \citet{Bethermin2017}, which synthesizes state-of-the-art constraints on the extragalactic cosmic infrared background (CIB), reproducing a wide variety of number count measurements in the mid-infrared (e.g., \textit{Spitzer}), far-infrared (e.g., \textit{Herschel}), and \submm{} (e.g., SCUBA2).
The extragalactic photometric confusion limits measured for a 50~m telescope assuming $q_{\rm phot}=5$ are summarized in Table~\ref{tab:conf_sdc}.
The very large diameter of AtLAST gives very low photometric confusion limits, of the order of a few tens of micro-Jansky, or even lower than the flux density range of our model in the two AtLAST bands with the highest angular resolution (i.e., 350 and 450~$\mu$m).  
These are about 100 times lower limits than the photometric confusion limits reached by current 10-m class single-dish \submm{} telescopes or even $>10\,000$ times lower than those of the \textit{Herschel} space telescope.
Only the LMT can match these low photometric confusion limits, but it is limited to a very small field of view ($\sim8'$) and lacks the essential sub-mm coverage.

The confusion defined by the so-called source density criterion \citep[SDC; e.g.][]{Dole2003} is driven by the fraction of sources lost in the detection process because the nearest neighbor whose flux is greater than $S_{\rm lim}^{\rm SDC}$ is too close to be separated.
Unlike the photometric confusion, the SDC confusion does not act as a background noise against which sources must be detected, but it affects the completeness of the catalog of sources of interest.  
It is therefore more relevant when studying the sources themselves and, for example, the SDC confusion limit set by distant galaxies does not directly affect galactic science (although galactic source catalogs are also affected by the same phenomena).
Following \citet{Dole2003}, the SDC confusion is defined using the probability $P$ given a source density $N_{\rm SDC}$ (Poisson distribution) of having the nearest source of flux $S_{\rm lim}^{\rm SDC}$ located at a distance less than $\theta_{\rm min}$,
\begin{equation}
\label{eq:conf SDC 1}
P( <\theta_{\rm min})= 1 - e^{-\pi N_{\rm SDC} \theta_{\rm min}^2},
\end{equation}
where $\theta_{\rm min}$ is usually defined in term of the FWHM of the beam as
\begin{equation}
\label{eq:conf SDC 2}
\theta_{\rm min} = k\,\theta_{\rm FWHM}.
\end{equation}
It follows that $N_{\rm SDC}$ is given by
\begin{equation}
\label{eq:conf SDC 3}
N_{\rm SDC} = - \frac{{\rm ln}\,(\,1 - P(<\theta_{\rm min})\,)}{\pi k^2 \theta_{\rm FWHM}^2}.
\end{equation}
By choosing values for $P(<\theta_{\rm min})$ and $k$, $N_{\rm SDC}$ can be calculated and then translated into $S_{\rm lim}^{\rm SDC}$ using a source number count model.
Although  $P(<\theta_{\rm min})$ depends on the desired completeness and $k$ on the exact source extraction method used, it is generally set to 10\% and 0.8, respectively \citep{Dole2003}.
This gives a $N_{\rm SDC}$ of 1 source per 16.7 beam area.
Using the galaxy population model of \citet{Bethermin2017}, we translated this source density into the extragalactic SDC confusion limits for a 50~m telescope (see Table~\ref{tab:conf_sdc}).
In most AtLAST bands, the extragalactic SDC confusion limits are greater than those deduced from the photometric criterion, with the transition between the photometric and source density criteria occurring only at 3~mm.
At these confusion limits, a very large fraction of the extragalactic CIB is nevertheless resolved into individual galaxies (see the third column of Table~\ref{tab:conf_sdc}), enabling unprecedented advances in our understanding of galaxy evolution (see, e.g., Section~\ref{SciTheme_DistantSurveys}).
We also note that these are rather conservative SDC confusion limits, as source extraction methods using prior information on source position (from, for example, complementary optical/near-infrared surveys) can separate sources down to $k=0.5$ \citep[e.g.,][i.e., a $N_{\rm SDC}$ of 1 source per 6.5 beam area]{Magnelli2009}.
This lowers the extragalactic SDC confusion limits by a factor $\sim3$ and leads to a transition between the photometric and source density criteria at 1.3~mm. 
In this more optimistic case, more than $80\%$ of the extragalactic CIB is resolved in the AtLAST bands shortwards of 1.3~mm.

Our calculations unambiguously demonstrate the ability of a 50~m telescope to achieve unrivalled low confusion limits, enabling deep galactic observations and resolving into individual galaxies most of the extragalactic CIB in the 0.35 to 1.3~mm range. 
As shown in Section~\ref{SciTheme_DistantSurveys}, such deep observations over a wide area of the sky ($>1\,000\,$deg$^2$) will in particular enable us to study the far-infrared/\submm{} emission of a large and comprehensive sample of typical star-forming galaxies (i.e. at $L^\star$) from $z=0$ to $z\sim5$.
By comparison, a 40~m telescope would have $\sim1.8$ $\times$ higher confusion limits, resolving in each band about 15\% less of the extragalactic CIB into individual galaxies, and allowing the study of typical star-forming galaxies only up to $z\sim 4$.

\begin{table}[]
   \centering
    \begin{tabular}{c|ccc}
        Wavelength & Photometric Confusion Limit  & SDC Confusion Limit  & CIB Resolved\\
        $\mu$m & $\mu$Jy & $\mu$Jy & \% \\
        \hline %
        350  & $<$1 & \textbf{56}  & 97 \\
        450  & $<$1 & \textbf{112} & 92 \\
        750  & 30   & \textbf{195} & 76 \\
        850  & 34   & \textbf{166} & 70 \\
        1100 & 44   & \textbf{141} & 61  \\
        1300 & 65   & \textbf{138} & 50 \\
        2000 & 57   & \textbf{68}  & 36 \\
        3000 & \textbf{39}   & 31  & 16  \\
    \end{tabular}
    \caption{Extragalactic confusion limits based on the photometric (assuming $q_{\rm phot}=5$) and source density (SDC; assuming $P(\theta_{\rm min}) = 0.1$ and $k=0.8$) criteria. The third column gives the fraction of the extragalactic CIB resolved into individual galaxies at the higher of these two limits (highlighted in bold). These calculations are based on the galaxy population model of \citet{Bethermin2017}.
    }
    \label{tab:conf_sdc}
\end{table}

\section{AtLAST Sensitivity Calculator and Simulator}
\label{app:senscalc}

To allow astronomers contributing to the science working groups the ability to understand whether their planned observations were realistic, and to ensure consistent calculations were being done across working groups, a sensitivity calculator and telescope simulator was developed.

Having a single source of truth, from which astronomers can derive integration times required to reach their desired sensitivity (or conversely, achieved sensitivity in a given on-source integration time), is vital for ensuring that astronomers are comparing like-for-like when deriving observing cases.

At such an early stage in the development of the telescope, when so many details (large and small) are still under constant review, the goal of the sensitivity calculator was to be as flexible as possible, even including the ability to modify the diameter of the telescope when running the calculator from the source code.

The sensitivity calculator was designed with simplicity and flexibility of use in mind. As such, users can interact with it either through a simplified web interface, with most telescope parameters fixed, or at the command line (or within a python instance) and configure any and all of the input parameters to the sensitivity calculation to their specification.  The former lowers the barriers to participation by removing the complexity in setting up the calculation, the latter allows expert users to explore the parameter space available to them through the calculation to better understand how varying a telescope property (like dish diameter) will affect their results.

It is through the command line (or python) interface that users can create simulated observations using an input \texttt{FITS} dataset (2D or 3D), with examples of how to do that, in the form of Jupyter Notebooks, provided within the sensitivity calculator source code.  With these tools, astronomers can simulate observing with an idealised telescope under various weather conditions with the caveats and assumptions provided below.

\subsection{Underlying Equations, Assumptions and Caveats}

As explained in the \href{https://atlast-sensitivity-calculator.readthedocs.io/en/latest/calculator_info/sensitivity.html}{sensitivity calculator documentation}, we used the system equivalent flux density (SEFD) of a telescope to derive the sensitivity achievable in a given integration time or vice versa:

\begin{equation}
                    \Delta S = \frac{SEFD}{\eta_{s}\sqrt{2\Delta \nu t}} \hspace{20pt}\textrm{or} \hspace{20pt}t  = \left(\frac{SEFD}{\Delta S \eta_s}\right)^2 \frac{1}{n_{pol}\Delta\nu} 
\end{equation}

\noindent where $\eta_s$ is the system efficiency, $n_{pol}$ is the number of polarisations, $\Delta \nu$ is the bandwidth under consideration (of a single channel for spectral line observations, or the full bandwidth for continuum) and the SEFD is defined as:

\begin{equation}
    SEFD = \frac{2kT_{sys}}{\eta_A A_g}
\end{equation}

\noindent where $k$ is the Boltzmann constant, $T_{sys}$ is the system temperature, $\eta_A$ is the aperture efficiency, and $A_g$ is the geometric dish area.  Further breakdowns of the equations used in the sensitivity calculator can be found in the online documentation that accompanies the code and web client.

\subsubsection{Assumptions}

With the telescope design still actively under construction, and the final efficiencies and transmission coefficients unknown, a number of assumptions about these parameters, as well as those relating to as yet to be defined instrumentation suites.

\begin{itemize}
    \item Telescope efficiencies such as system, dish, forward, spillover, illumination, polarisation and blocking were assumed based on parameters from other similar facilities
    \item Atmospheric and sky transmissions and temperatures were based on measurements of conditions on the Chajnantor Plateau
    \item No specific instrument efficiencies were used because they depend on evolving underlying technology 
    \item Receiver temperatures were assumed to be near their fundamental quantum limit at $T_{rx} = 5h\nu / k$
\end{itemize}

\subsubsection{Caveats}

Because there are a number of unknowns with respect to the final operational model, instrument suite, and calibration strategies of the observatory, the following caveats apply to the sensitivity calculations: 
\begin{itemize}
    \item Only on-source integration times are provided. This is because the operational model, calibration strategies and mapping patterns/speeds are not yet known for the observatory.
    \item No pixel spacing information is inherent in the calculations because that is highly dependent on how the future instrumentation will be implemented.  For the simulations, a pixel size consistent with diffraction limited seeing is applied to the input image or spectrum. 
\end{itemize}

\subsection{Implementation}

The code was implemented with a python backend. For the web client, a javascript front end has been implemented. The source code is available through \href{https://github.com/ukatc/AtLAST_sensitivity_calculator}{github} with the web front end can be accessed through the \href{https://www.atlast.uio.no/sensitivity-calculator/}{AtLAST website}.  In both instances, best software engineering practices were employed, with an emphasis on modularity, testing and documentation to allow for future development of the calculator as the project evolves.

\subsubsection{Source code}

\texttt{Python} was chosen for backend development because of its prevalence within the astronomical community: enabling interested astronomers to modify parameters of the simulation with relative ease.  The backend code is highly modular to allow for as much code reuse as possible and to simplify updates.  A description of the public and REST APIs as well as UML class diagrams are available as part of the documentation.

All parameters relevant to the sensitivity calculation have been abstracted so that they can be modified by the user using dedicated input files, or specified at runtime at the command line, or within a python instance. This latter mode allows astronomers to provide the calculators with a set of telescope properties which vary (i.e. telescope diameter ranging from 40 to 60 m) in a loop to generate tables of output sensitivities or time estimates.  This configurability is useful as the telescope properties (and eventually, the instrument properties) are refined.  Semantic validation is performed using 

All of the code is subject to unit and functional testing to ensure that as code is updated and modified, the underlying functionality changes in expected ways (i.e. if increasing calculation efficiency, the output results do not change). For maintainability, the code is linted and adheres to PEP8 and Black formatting principles.

Extensive \texttt{readthedocs} documentation was created for 1) understanding the individual code components 2) for installing the code in its own \texttt{conda}, \texttt{venv} or \texttt{poetry} environment and 3) user guides for using the calculator and web client.

\subsubsection{Web Client}

\texttt{Javascript} was chosen for the web client because of its ease of use for simple projects and is consistent with other calculators being developed by other facilities. As with the backend, best software engineering principles were used, and a minimal set of parameters are passed between the front and back ends of the project to minimise traffic and lag time.  Client side validation is performed before any properties are passed to the backend for processing.

\subsection{Imaging and Spectroscopic simulator}

Included in the source code repository are two Jupyter Notebooks which show how the sensitivity calculator can be used to derive telescope simulations. These notebooks take as input the set of observing properties required for the calculation (the same ones as can be found in the web version of the calculator) and an input \texttt{FITS} file in either two or three dimensions, depending on the notebook being used. The notebooks then use the sensitivity calculator to derive the  noise levels to apply to the input image based on the telescope parameters and on-source integration times specified. These noise levels (as Gaussian noise) are then applied to a new version of the input image which has been convolved with a Gaussian beam corresponding to the diffraction limit of a 50m telescope at the input wavelength.

\subsection{Related simulators}

The \texttt{Maria} simulator \citep{vanMarrewijk2024b} was created alongside the sensitivity calculator described above to simulate the time variable atmosphere from various sites around the world used for (sub-)mm observations. It can be used to optimise observing strategies. Its performance has been tested against the actual performance of the \texttt{Mustang-2} instrument on the Green Bank Telescope, and predictions have been made for what can be expected for large scale source recovery with AtLAST.  \texttt{Maria} creates timeseries data which can then be transformed into sky maps using data reduction software. The timestream data can be exported into \texttt{FITS} or \texttt{HDF5} file formats to allow for ingest into the relevant packages for the telescope being simulated. 

\section{Acronyms}
\label{app:acronyms}

\begin{center}
\begin{longtable}{|l|l|}
\hline
Acronym & Meaning \\
\hline
\endhead
ACA	& Atacama Compact Array (Morita Array, 7-metre component of ALMA)\\
\href{https://act.princeton.edu/}{ACT}	& Atacama Cosmology Telescope\\
AGN & Active Galactic Nucleus \\
\href{https://almaobservatory.org}{ALMA} &  Atacama Large Millimeter/submillimeter Array\\
\href{http://www.apex-telescope.org/}{APEX} & Atacama Pathfinder Experiment (APEX)\\
\href{https://atlast-telescope.org}{AtLAST} &  Atacama Large Aperture Submillimetre Telescope  \\
\href{https://www.nao.ac.jp/en/research/telescope/aste.html}{ASTE} & Atacama Submillimetre Telescope Experiment\\
\href{http://bicepkeck.org/}{BICEP} & Background Imaging of Cosmic Extragalactic Polarization\\
CGM & Circumgalactic medium \\
CIB & Cosmic infrared background \\
CMB & Cosmic microwave background \\
\href{https://cmb-s4.org/}{CMB-S4}	& The US Dept of Energy backed ``Stage IV'' proposed CMB experiment\\
\href{http://cso.caltech.edu/}{CSO} & Caltech Submm Observatory \\
\href{https://eventhorizontelescope.org/}{EHT} & Event Horizon Telescope\\
FoV	& Field of View\\
FYST & Fred Young Submm Telescope (formerly CCAT)\\
GMC & Giant Molecular Cloud \\
GRB & Gamma Ray Burst \\
ICM & Intracluster Medium \\
IFU & Integral Field Unit\\
\href{https://www.iram-institute.org}{IRAM} &	Institut de Radioastronomie Millimétrique\\
ISM & Interstellar Medium \\
\href{https://www.eaobservatory.org/jcmt/}{JCMT} & James Clerk Maxwell Telescope\\
JWST &	James Webb Space Telescope\\
KAO & Kuiper Airborne Observatory \\
KID & Kinetic Inductance Detector\\
LAS & Largest Angular Scale \\
\href{http://www.lmtgtm.org/}{LMT}	& Large Millimeter Telescope \\
\href{https://www.lsst.org/}{LSST} & The Vera C. Rubin Observatory, formerly ``Large Synoptic Survey Telescope''\\
MOS & Multi-Object Spectragraph \\
MRS & Maximum Resolvable Scale \\
\href{https://ngvla.nrao.edu/}{ngVLA} & The next-generation Very Large Array\\
\href{https://www.iram-institute.org/EN/noema-project.php?ContentID=9&rub=9&srub=0&ssrub=0&sssrub=0}{NOEMA} & NOrthern Extended Millimeter Array\\
\href{https://skao.int}{SKAO} & Square Kilometer Array Observatory \\
\href{https://www.cfa.harvard.edu/sma/}{SMA} & Submillimeter Array\\
\href{https://simonsobservatory.org/}{SO} & Simons Observatory\\
\href{http://www.ioa.s.u-tokyo.ac.jp/TAO/en/}{TAO} & (University of) Tokyo Atacama Observatory \\
WHIM & Warm/Hot Ionised Medium \\
\hline
\end{longtable}
\label{tab:acronyms}
\end{center}

\section{Author affiliations}
\label{app:affil}

$^1$UK Astronomy Technology Centre, Royal Observatory Edinburgh, Blackford Hill, Edinburgh EH9 3HJ, UK\\
$^2$Institute of Theoretical Astrophysics, University of Oslo, PO Box 1029, Blindern 0315, Oslo, Norway\\
$^3$European Southern Observatory (ESO), Karl-Schwarzschild-Strasse 2, Garching 85748, Germany\\
$^4$Astrochemistry Laboratory, Code 691, NASA Goddard Space Flight Center, Greenbelt, MD 20771, USA\\
$^5$Laboratoire Lagrange, Université Côte d'Azur, Observatoire de la Côte d'Azur, CNRS, Blvd de l'Observatoire, CS 34229, 06304 Nice cedex 4, France\\
$^6$Astronomy Unit, Department of Physics, University of Trieste, via Tiepolo 11, Trieste 34131, Italy\\
$^{7}$INAF -- Osservatorio Astronomico di Trieste, via Tiepolo 11, Trieste 34131, Italy\\
$^{8}$IFPU -- Institute for Fundamental Physics of the Universe, Via Beirut 2, 34014 Trieste, Italy\\
$^{9}$NRC Herzberg Astronomy and Astrophysics, 5071 West Saanich Rd, Victoria, BC, V9E 2E7, Canada\\
$^{10}$Department of Physics and Astronomy, University of Victoria, Victoria, BC, V8P 5C2, Canada\\
$^{11}$Cosmic Dawn Center (DAWN), Denmark\\
$^{12}$DTU-Space, Technical University of Denmark, Elektrovej 327, DK2800 Kgs. Lyngby, Denmark\\
$^{13}$Max-Planck-Institut f\"{u}r extraterrestrische Physik, Giessenbachstrasse 1 Garching, Bayern, D-85748, Germany\\
$^{14}$Purple Mountain Observatory, Chinese Academy of Sciences, 10 Yuanhua Road, Nanjing 210008, China\\
$^{15}$Department of Physics and Astronomy, University of Pennsylvania, 209 South 33rd Street, Philadelphia, PA, 19104, USA\\
$^{16}$Department of Physics and Astronomy, University College London, Gower Street, London WC1E 6BT, UK\\
$^{17}$Max-Planck-Institut f\"ur Radioastronomie (MPIfR), Auf dem H\"ugel 69, D-53121 Bonn, Germany\\
$^{18}$School of Physics \& Astronomy, Cardiff University, The Parade, Cardiff CF24 3AA, UK\\
$^{19}$Division of Geological and Planetary Sciences, California Institute of Technology, Pasadena, CA 91125, USA.\\
$^{20}$Rosseland Centre for Solar Physics, Institute of Theoretical Astrophysics, University of Oslo, Postboks 1029 Blindern, N-0315 Oslo, Norway\\
$^{21}$Haystack Observatory, Massachusetts Institute of Technology, 99 Millstone Road, Westford, MA 01886, USA\\
$^{22}$National Astronomical Observatory of Japan, 2-21-1 Osawa, Mitaka, Tokyo 181-8588, Japan\\
$^{23}$Black Hole Initiative, Center for Astrophysics $\mid$ Harvard \& Smithsonian, 20 Garden Street, Cambridge, MA 02138, USA\\
$^{24}$INAF Osservatorio Astronomico di Brera, via Brera 28, 20121 Milano, Italy\\
$^{25}$National Astronomical Observatory of Japan, Osawa 2-21-1, Mitaka, Tokyo 181-8588, Japan\\
$^{26}$Department of Space, Earth, \& Environment, Chalmers University of Technology, Chalmersplatsen 4 412 96 Gothenburg, Sweden\\
$^{27}$Université de Strasbourg, CNRS, Observatoire astronomique de Strasbourg, UMR 7550, F-67000 Strasbourg, France \\
$^{28}$Academia Sinica Institute of Astronomy and Astrophysics, 645 N. A'ohoku Place, Hilo, HI 96720, USA\\
$^{29}$Hvar Observatory, Faculty of Geodesy, University of Zagreb, Zagreb, Croatia\\
$^{30}$Academia Sinica Institute of Astronomy and Astrophysics (ASIAA), No. 1, Sec. 4, Roosevelt Road, Taipei 106216, Taiwan\\
$^{31}$International Centre for Radio Astronomy Research (ICRAR), The University of Western Australia, M468, 35 Stirling Highway, Crawley, WA 6009, Australia\\
$^{32}$Armagh Observatory and Planetarium, College Hill, Armagh, BT61 9DB, UK\\
$^{33}$INAF, Osservatorio di Astrofisica e Scienza dello Spazio, via Piero Gobetti 93/3, 40129 Bologna, Italy\\
$^{34}$INFN, Sezione di Bologna, viale Berti Pichat 6/2, 40127 Bologna, Italy\\
$^{35}$Instituto de Astrofisica de Canarias (IAC), E-38205 La Laguna, Tenerife, Spain\\
$^{36}$Universidad de La Laguna, Dpto. Astrofısica, E-38206 La Laguna, Tenerife, Spain\\
$^{37}$School of Physics, Trinity College Dublin, College Green, Dublin 2, Ireland\\
$^{38}$Université Paris-Saclay, Université Paris Cité, CEA, CNRS, AIM, 91191, Gif-sur-Yvette, France\\
$^{39}$Departamento de Física de la Tierra y Astrofísica e Instituto de Física de Partículas y del Cosmos (IPARCOS). Universidad Complutense de Madrid, Av. Complutense, s/n, 28040 Madrid, Spain\\
$^{40}$Department of Astronomy, Cornell University, Ithaca, NY 14853, USA; Department of Physics, Cornell University, Ithaca, NY 14853, USA\\
$^{41}$Planetary Systems Laboratory, NASA Goddard Space Flight Center, Greenbelt, MD, 20771, USA\\
$^{42}$Departments of Astronomy and of Earth and Planetary Science, University of California Berkeley, Berkeley, California, 94720, USA\\
$^{43}$Victoria University of Wellington, Wellington, New Zealand\\
$^{44}$Physics and Astronomy, University of Southampton, Highfield, Southampton, SO17 1BJ, UK\\
$^{45}$DARK, Niels Bohr Institute, University of Copenhagen, Jagtvej 155, Copenhagen N, 2200, Denmark\\
$^{46}$Max Planck Institute for Extraterrestrial Physics, Garching bei München, 85748, Germany\\
$^{47}$Jodrell Bank Centre for Astrophysics, University of Manchester, Manchester, UK\\
$^{48}$Leiden Observatory, Leiden University, P.O. Box 9513, 2300 RA Leiden, the Netherlands\\
$^{49}$Faculty of Electrical Engineering, Mathematics and Computer Science, Delft University of Technology, Mekelweg 4, 2628 CD Delft, The Netherlands\\
$^{50}$SRON – Netherlands Institute for Space Research, Niels Bohrweg 4, 2333 CA Leiden, The Netherlands\\
$^{51}$School of Mathematics, Statistics and Physics, Newcastle University, Newcastle upon Tyne, NE1 7RU, UK\\
$^{52}$Max Planck Institute for Astronomy, Koenigstuhl 17, 69117 Heidelberg, Germany\\
$^{53}$Instituto de Radioastronomía y Astrofísica (IRyA), Universidad Nacional Autónoma de México (UNAM), Antigua Carretera a Pátzcuaro, 8701, Ex-Hda. San José de la Huerta, Morelia, Michoacán, 58089, México\\
$^{54}$INAF-Osservatorio Astrofisico di Arcetri, Largo Enrico Fermi 5, I-50125 Firenze, Italy\\
$^{55}$National Radio Astronomy Observatory, Socorro, NM, 87801, USA\\
$^{56}$University of Milan Bicocca, Piazza della Scienza 3, Milan, Italy\\
$^{57}$Astrophysics Department, Instituto Nacional de Astrofísica, Óptica y Electrónica, Luis E. Erro 1, Tonantzintla, Puebla, C.P. 72840, México\\
$^{58}$National Radio Astronomy Observatory, 520 Edgemont Road, Charlottesville, VA 22903, USA\\
$^{59}$School of Physics and Astronomy, University of Leicester, Leicester, LE1 7RH, UK\\
$^{60}$Center for Solar-Terrestrial Research, New Jersey Institute of Technology, Newark, NJ, 07102, USA\\
$^{61}$Astronomical Institute, Czech Academy of Sciences, 25165, Ondřejov, Czech Republic\\
$^{62}$Department of Astrophysics, University of Vienna, Türkenschanzstrasse 17, A-1180 Vienna, Austria\\
$^{63}$Nicolaus Copernicus Astronomical Center, Polish Academy of Sciences, Rabiańska 8, 87-100, Toruń, Poland\\
$^{64}$Division of Geological and Planetary Sciences, California Institute of Technology, Pasadena, California, 91125, USA\\
$^{65}$School of Physics \& Astronomy, University of Glasgow, Glasgow, G12 8QQ, UK\\
$^{66}$National Taiwan Normal University, Taipei City, 116, Taiwan\\
$^{67}$Kavli Institute for Particle Astrophysics \& Cosmology (KIPAC), Stanford University, Stanford, CA 94305, USA\\
$^{68}$Aix Marseille Univ, CNRS, CNES, LAM, Marseille, France\\
$^{69}$NASA Goddard Space Flight Center, Greenbelt, MD 20771, USA\\
$^{70}$The Catholic University of America, Washington, DC 20064, USA\\
$^{71}$LESIA, Observatoire de Paris, Université PSL, CNRS, Sorbonne Université, Université Paris Cité, 5 place Jules Janssen, 92195 Meudon, France\\
$^{72}$Central  Astronomical Observatory at Pulkovo of Russian Academy of Sciences, St. Petersburg, 196140, Russia\\
$^{73}$Ioffe Institute, Polytekhnicheskaya, 26, St. Petersburg, 194021, Russia\\
$^{74}$INAF Osservatorio Astronomico di Cagliari, Via della Scienza 5, 09047 Selargius (CA), Italy\\
$^{75}$Instituto de Estudios Astrof\'isicos, Facultad de Ingenier\'ia y Ciencias, Universidad Diego Portales, Av. Ej\'ercito Libertador 441, Santiago, Chile\\
$^{76}$Kavli Institute for Astronomy and Astrophysics, Peking University, Beijing 100871, China\\
$^{77}$School of Physics and Astronomy, University of Leeds, Leeds LS2 9JT, UK\\
$^{78}$Institut de Ci\`encies de l'Espai (ICE, CSIC), Campus UAB, Carrer de Can Magrans s/n, 08193, Bellaterra (Barcelona), Spain\\
$^{79}$Institut d'Estudis Espacials de Catalunya (IEEC), 08860, Castelldefels (Barcelona), Spain\\
$^{80}$National Astronomical Observatory of Japan, Mitaka, Tokyo, 181-8588, Japan\\
$^{81}$Graduate University of Advanced Studies, SOKENDAI, Mitaka, Tokyo 181-8588, Japan\\
$^{82}$IAPS-INAF, Rome, I-00133, Italy\\
$^{83}$Dipartimento di Fisica e Astronomia 'Augusto Righi', Universit\`a degli Studi di Bologna, Via Gobetti 93/2, I-40129 Bologna, Italy\\
$^{84}$Air Force Research Laboratory, Space Vehicles Directorate, Kirtland AFB, NM 87123, USA\\
\begingroup
\bibliographystyle{apj_mod}
\setlength{\parskip}{0pt}
\setlength{\bibsep}{0pt}
\bibliography{references,refs/galaxy,refs/cgm,refs/solar,refs/ssystem,refs/vlbi,refs/atlast,refs/sz,refs/transients,refs/distant,refs/nearbygal}
\endgroup
\end{document}